\documentclass{article}
\usepackage{graphicx} % Required for inserting images
\usepackage{todonotes}
\usepackage{lscape}
\usepackage{authblk}
\usepackage[colorlinks,citecolor=blue,urlcolor=blue,bookmarks=false,hypertexnames=true]{hyperref} 
\usepackage{xcolor}
\usepackage{adjustbox}
\usepackage{threeparttable}
\usepackage{subcaption}
\usepackage{todonotes}
\usepackage[utf8]{inputenc}
\usepackage[T1]{fontenc}
\usepackage[bottom]{footmisc}
\usepackage{lmodern} 
\usepackage{amsmath}
\usepackage{comment}
\usepackage{hyperref}
\usepackage{caption}
\usepackage{subcaption}
\usepackage{longtable}
\usepackage[style=apa]{biblatex}
\usepackage{enumitem}
\usepackage{hyperref}
\usepackage{subcaption}
\usepackage{upquote} % Fixes symbol rendering
\usepackage{listings}

\usepackage[a4paper]{geometry}
\title{pyRMV: Reusable Cross-Model Validation for Computational Science}
\author[1]{Hugo Dictus}
\author[1]{Eduard Subert}
\author[1]{Armando Romani}
\author[1]{Henry Markram}
\affil[1]{Blue Brain Project, École polytechnique f\'ed\'erale de Lausanne (EPFL), Campus Biotech, Geneva, Switzerland}

\begin{document}
\maketitle

\lstdefinestyle{mystyle}{
    backgroundcolor=\color{white},   
    commentstyle=\color{green},
    keywordstyle=\color{magenta},
    numberstyle=\tiny\color{gray},
    basicstyle=\linespread{0.6}\small \ttfamily,
    stringstyle=\color{purple},
    columns=fullflexible,
    showstringspaces=false,
    % basicstyle=\ttfamily\footnotesize,
    % breakatwhitespace=false,     
    % breaklines=true,                 
    % captionpos=b,                    
    % keepspaces=true,                 
    % numbers=left,                    
    % numbersep=5pt,                  
    % showspaces=false,                
    % showstringspaces=false,
    % showtabs=false,                  
    % tabsize=2
    linewidth=\textwidth
}
\lstset{style=mystyle}

% journal options:
% springer Neuroinformatics https://link.springer.com/journal/12021/submission-guidelines#Instructions%20for%20Authors_Types%20of%20papers
% frontiers in neuroinformatics https://www.frontiersin.org/journals/neuroinformatics/for-authors/article-types
% Nature Scientific Data https://www.nature.com/sdata/submission-guidelines#sec-1
% Plos Computational Biology https://journals.plos.org/ploscompbiol/s/submission-guidelines
% elife https://elifesciences.org/subjects/computational-systems-biology

% TODO: improve the name
% TODO: ease off of "most important problem"
% TODO: Put properties as interface more forward - it forms a strong conceptual core for both strengths and limitations
% - in overcoming the limitations, you can refer back to it
% - and shorten property definition ("to paraphrase humphreys")
% TODO: include strengths and limitaitons in table
% TODO: cite some articles bemoaning the lack of standards in other fields

\newpage
\section*{Abstract}
We present a model-agnostic approach to the validation of computational models, and a python library containing a suite of validations for mouse primary visual cortex models.
By viewing the model not as a stand-in for the target system, but instead as a generator of predicted system properties, this approach allows validations to easily generalize to many different models, while also making them comparatively simple to write.
Several limitations of this approach are outlined alongside possible means of addressing them.
This presents an important step in addressing what we consider the most important problem in computational science: the standardization of model validation.

\newpage

\section{Introduction}

Evaluating how accurately a computational model represents its target system is a critical step in any computational study. This step is called validation.
It is generally infeasible to validate models against all of the relevant data, and this feasibility only decreases as the scope, scale, and detail of simulations increases.
This is largely because new models must implement validation procedures from scratch.
Different models differ in their implementation and representation of the target system - precluding reuse of previous validation code.
Addressing this is a necessity for building, refining, and replacing increasingly realistic models of highly complex systems.
This challenge is increasingly recognized in numerous computational fields \parencite{tatka_adapting_2023, fahsbender_benchmarking_2025, chudziak_studying_2025,schrimpf_integrative_2020,larooij_large_2025}. 
We present here pyRMV, a software framework for the creation and re-use of validation code across differing models, as well as the conceptual advances which inform its design.

% especially important for combining fields, or fields where laws are not well known
% especially important for large-scale detailed simulations, which have many, many respects in which they want to be realistic
% or for simulations of rules with far-ranging consequences, such as plasticity rules. You want to check these against as wide a range of empirical constraints as possible.

In neuroscience, software engineering techniques to facilitate code reuse have been applied to validations across different computational models through the SciUnit package \parencite{omar_collaborative_2014,gerkin_neuronunit_2019,gutzen_evaluating_2019,appukuttan_software_2022} and the Brain-Score platform \parencite{schrimpf_integrative_2020}, with similar techniques used in modeling competitions  \parencite{willeke_sensorium_2022,turishcheva_dynamic_2024}. The key technique used is to create an interface representing the model which assumes as little as possible about its implementation, an application of abstraction by specification \parencite{liskov_abstraction_1986}.

Existing approaches generally usually involve a validation performing a series of operations on a model object to mimic the experimental procedure. The model is then expected to provide methods to simulate those operations. This approach therefore treats the measurements of an experiment as the outcomes of a set of operations rather than as observations of an underlying property, consistent with the philosophical tradition of operationalism \parencite{bridgman_logic_1927}. We will call this approach to validation operationalist validation. An operationalist perspective places severe limitations in terms of the ability to scale, generalize, or simplify validation. We contrast it with an approach in which the measurements reported by and experiment and predicted by a model are both treated as though they correspond to an underlying property of the target system. This is more consistent with the perspective of property-cluster realism \parencite{humphreys_extending_2004}, and we consequently call this approach 'realist'.
We believe this distinction is essential to understanding and solving the challenges faced in systematic model validation. To further this goal we implemented a software framework that enforces a realist approach to validation. 

PyRMV is a python package intended to make it easier to write validations and reuse them across models with arbitrary implementations or means of representing their target system. Models recieve some data on the basis of which to make predictions, and based on these return their predictions of the properties of the system under study. This allows models with fundamentally differing conceptual underpinnings to be directly compared. In this article we explain how pyRMV works and apply it to a mouse primary visual cortex model. In the process we evaluate its strengths and limitations and how these relate to the operationalist-realist distinction, and how they clarify unviversal challenges in systematic validation.

%Key to all of these techniques is the construction of an abstraction \cite{liskov_abstraction_1986} to represent the model. That is an interface constructed to represent the model, in which the model's details are ignored (abstracted away) and only the relevant features are present. The validation is written to use this interface, and thereby does not depend on the missing details. Models which are similar enough to satisfy the interface can therefore be used with the validation regardless of how they implement it.

\subsection{Existing approaches and the disadvantages of operationalism}

The SciUnit framework \parencite{omar_collaborative_2014} and associated neuroscience-specific libraries \parencite{gerkin_neuronunit_2019,gutzen_evaluating_2019,saray_hippounit_2021,appukuttan_software_2022} have been developed to validate models of neurons and brain regions. They use classes called Capabilities to specify which methods a model class must implement to support a validation, and the model class must subclass the Capability to run the validations that require it. This inheritance-based approach leads to a fairly complex and rigid hierarchy of classes that adds cognitive load and lines of code to any effort to construct a validation or validate a model.
Additionally, it is difficult to effectively reuse capabilities across validations and especially across different brain regions or modeling traditions.
This is due to the fact that capabilities may be  specific to the experiment and model they were first designed for, and because the framework is fragmented across many repositories (an issue made more complex by the dependencies introduced by inheritance).

The Brain-Score \parencite{schrimpf_integrative_2020} platform and  sensorium competition \parencite{turishcheva_dynamic_2024,willeke_sensorium_2022} side-step many of these issues by constraining their scopes upfront: focusing on specific types of models and experiments. This allows them to provide specific support and interfaces which make the submission of models easier. This is beneficial with the objective of evaluating many models on a body of similar experiments, but when the objective is to comprehensively validate a mechanistic model, as in simulation neuroscience applications \parencite{markram_reconstruction_2015,billeh_systematic_2020}, the wide variety of experiments makes such specialization impractical. Furthermore, this specialization limits the number of ways in which a model's match with the target system can be evaluated, and makes comparison with models that have very different underpinnings difficult.

A common feature of previous frameworks that they usually include a model of the experimental procedure within the validation\footnote{Strictly speaking, SciUnit does not explicitly require a representation of the experiment. Early example code for SciUnit in particular had the model provide predicted quantities. However, existing validations constructed with it most often define operationalst interfaces}. While this can save modelers the effort of creating such models themselves, it also places restrictions on them. Some models may not explicitly represent the elements present in the experimental procedure. For example, some validations in NetworkUnit that use firing rates require models to produce spike trains, and calculate the firing rates from these. This precludes the application of this validation to models which predict firing rates without explicitly representing spike times.
Conversely, other models may have a more comprehensive representation than anticipated by the validation. In the Brain-Score framework, linear regression is used to fit model firing rates to fMRI BOLD responses, which prevents models that include metabolic or vasculature modeling from making use thereof. As a result of this it is difficult to compare models with fundamentally differing approaches to representing the target system through these frameworks, while this would arguably be the greatest asset of generalized, systematic validation.

While the inclusion of a model of the experiment in the validation can sometimes save modelers effort, when different experimental methods measure the same underlying property it can instead increase the effort required for validation. For example, \cite{cossell_functional_2015} and \cite{lee_anatomy_2016} both measure connectivity between neurons, but the former does so with pair recordings and the latter with electron-microscopy. In this case, modeling the two experiments would require two completely different codebases. If the model instead simply provides predictions of the measured properties, it does not need additional code for the two different validations.

Additionally, a validation usually cannot describe the experiment exhaustively. Consider a data source like the MICrONs dataset \parencite{consortium_functional_2021}, which involved recording, slicing, and machine-assisted reconstruction and labeling. Modeling this experimental procedure would be prohibitively complex, and only be possible with models of the highest level of detail. Any validation must simplify such an exprimental procedure, and in doing so introduces theory-laden assumptions which any given model may dispute. Placing the modeling and interperetation of the experiment on the model side of the validation/model split would allow different interperetations of the experiment to compete in their capacity to explain it.

\subsection{Realist Model Validation}

Our conceptual approach is based on the perspective that the relationship between a model and its target system is mediated by the properties of the target system which the model represents \parencite{humphreys_extending_2004,nguyen_how_2016,graebner_how_2018}. Rather than a correspondence of model entities to objects in the system, or of processes in the model to processes in the target system, it is the imputation of model variables onto the target system by explication, representation, and prediction that defines its representation. This flexible conceptualization permits models with widely varying degrees of abstraction and different proposed underlying structures to be assessed by the same metrics, at least in principle. At length we will highlight some ways in which this flexibility is not necessarily present in practice, and how our approach can be improved to better preserve it.

% \begin{table}
%   \centering
%   \caption{Contrasting operationalist and realist validation}
%   \label{tab:operationalist-vs-realist}
%   \begin{tabular}{|c|c|c|}
%     \hline
%     ~ & Operationalist validation & realist validation \\ \hline \hline
%     The model must & simulate experimental procedures & predict measured properties \\ \hline
%     Code is & specific to experiment & reused across experiments \\ \hline
%     mandates & Specific model of experiment & no model of experiments \\ \hline
%     compares & similar models from same paradigm & any models which can represent the same properties \\ \hline
    
%   \end{tabular}
% \end{table}

% TODO: does this really add anything?
For our purposes. properties are inferred variables which act to unify different observations. For example, the temperature of a room is a property that explains the co-incidence of the sensation of feeling warm in that room, the level in a mercury thermometer, and various states of other observables. Properties can emerge from more basic properties, such as temperature emerging from the kinetic energy of many individual molecules. 

It can be argued that since properties are inferred, they belong to the model rather than to the validation; different models may posit different properties to explain the same observation. 
In practice however, theory and experiment are always holistically intertwined, because the raw observations are only available to the experimenter that observed them. Experimental results report inferred properties, which thereby serve as the interface between one experiment and another, as well as between experiments and models.
% This places a constraint on the generalizability of validations: there must be a basic level of agreement between model and experiment regarding the existence of the measured property.

% This constraint plays a role in the limitations we will identify and discuss.
% In an operationalist framework, you may simulate the procedure by which an experiment infers a property which does not exist within a model's conceptual framework. For example, modeling the measuring process by which an experiment erroneously infers a quantity of lost phlogiston, with an oxygen-based combustion model.
% Strategies 

%\subsection{Theoretical background : the DEKI framework}
%Consider not writing this section. It might be unneccessary waffle. 

\section{Methods}

\subsection{Design goals}

There are a number of different priorities which must be juggled in designing a validation framework. We outline our conception of them below.

\begin{enumerate}
    \item The effort required to support a model should be small, and \emph{must} scale sub-linearly with the number of validations
    \item the effort required to implement a new validation in the framework and run it on your model must be less than the effort to write and run the validation without the framework
    \item interfaces must be stable, such that modifying a validation seldom breaks compatibility with previous models 
    \item validations should impose as few restrictions as possible on the implementation or type of model they support
\end{enumerate}

Writing a standalone script will always be simpler than designing a reusable program which does the same job. Therefore, to satisfy item 2, the implementer must be able to easily re-use methods previously created for other validations. As someone implementing a new validation is likely doing so because they wish to run it on their model, the reusability of code in both the model and validation is important.

\subsection{Approach}

In pyRMV the model is a python object with methods corresponding to the measurable properties of the target system. These methods accept a dataframe containing the variables which act to further specify which properties are measured, such as the region of the brain from which to sample the property or the conditions under which to measure it. This defines the interface between model and validation. 

The experimental data, the experimental variables, and model predictions are all dataframes using a standardized terminology (\verb+.terminology+ module) and following tidy data conventions \parencite{wickham_tidy_2014}. 

Each column in the dataframe corresponds to a variable, defined as an instance of the Term class, which is itself a subclass of a string. The Term couples the name of the variable to a description which informally defines it and describes what values it can take. If it is a measurable variable, the name of the method which measures it is also defined. Currently, no formal specification is applied to the column, although we are considering this as a future improvement.

\begin{minipage}{\linewidth}
\begin{lstlisting}[language=Python]
FIRING_RATE = Term(
    "firing rate (Hz)",
    description=(
        "Rate of neuronal firing during a time window,"
        " spikes/second"
    ),
    measurement_method="firing_rate",
)
\end{lstlisting}
\end{minipage}

\begin{figure}
\begin{subfigure}[b]{0.45\textwidth}
   \includegraphics[width=\textwidth]{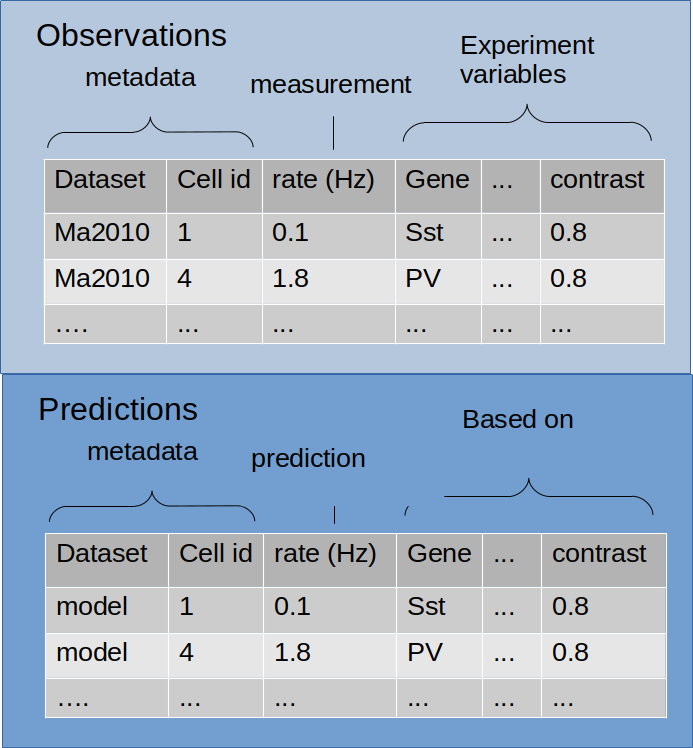}
 \end{subfigure}
 \begin{subfigure}[b]{0.45\textwidth}
   \includegraphics[width=\textwidth]{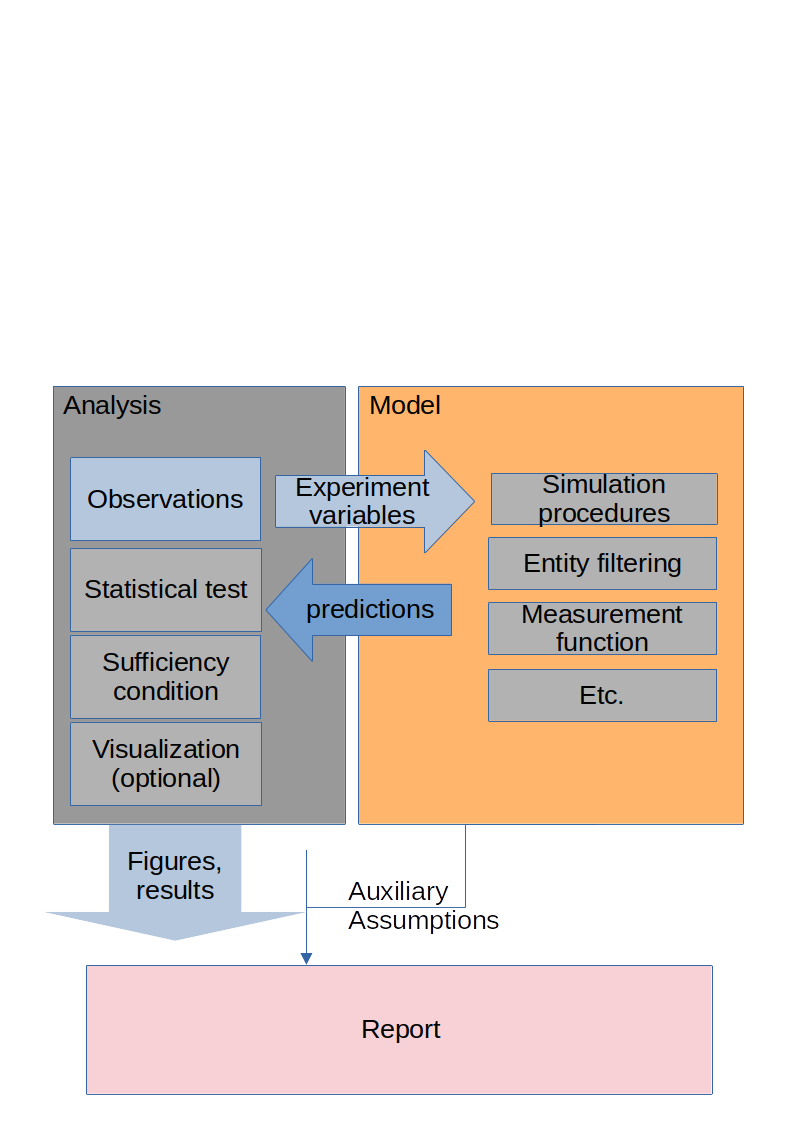}
 \end{subfigure}
   \caption{a: The same format is used for model and experimental data - the predictions of the model are based on the experimental variables extracted from the observations dataframe. b: Diagram representing framework operation. The validation consists of a number of components that can be replaced and reused as needed. Note that the model class can be constructed in any way, the internal structure is just to illustrate what sorts of tasks may be handled by the model}
\end{figure}

\begin{minipage}{\linewidth}
\begin{lstlisting}[language=Python]
# for illustrative purposes, not actual code
...
    exp_vars = self._extract_experimental_vars(self.observations, measurement)
    prediction = getattr(model, measurement.measurement_method)(exp_vars)
...
\end{lstlisting}
\end{minipage}

\begin{minipage}{\linewidth}
\begin{lstlisting}[language=Python]
# for illustrative purposes, not actual code
class MyModel:

    def orientation_selectivity(self, exp_vars):
       ...

    def firing_rate(self, exp_vars):
        ...

    def connection_probability(self, exp_vars):
        ...

    def cell_density(self, exp_vars):
        ...

    # private utility methods which handle vars
    def _filter_cells(self, cell_vars):
       ...
       
    def _filter_pairs(self, pair_vars):
       ...

    def _get_simulation(self, simulation_vars):
       ...
\end{lstlisting}
\end{minipage}

This differs from previous frameworks in that operations of the experiment are not represented in the validation, only its inferred measurements. For example, in assessing the firing rate of a neuron model in response to a stimulus, the validation may require the methods \verb+inject_square_current+ and \verb+get_voltage_traces+ to mimic the experimental procedure of an electrophysiology experiment. Our approach would require the method \verb+firing_rate+ and provide information about the stimulus current injected as an argument to this method. 

%For example, in visual neuroscience orientation selectivity refers to the tendency of neurons to respond more strongly to visual contours of a particular orientation, and this can be quantified with an Orientation Selectivity Index (OSI).
%An experiment \cite{siegle_survey_2021} provides orientation selectivity indices measured from various neurons in response to a given set of oriented visual stimuli. The neurons classes are defined in terms of the region in which they are situated, the layer of that region in which they are situated (generally based on depth), and whether spiking class (fast-spiking or regular-spiking, based on the width of their action potentials). These additional data specify exactly which properties of the target system were measured. 

% - note other measurements of orientation selectivity may not have same parameters

%When examined more closely, this particular example is also highly revealing of the limitations of our declarative approach, which we will expand on in \ref{sec:future-work}.

The format of the data passed between model and validation is constrained by a standardized terminology referring to column names and data types. This has some advantages over imperative, operationalist methods.

Firstly, when two validations assess the same property but compare to experiments with different experimental procedures the same code can be used for both. This reduces the number of methods needed on the model class, satisfying design goal 1. It also reduces the amount of code that needs to be written for a new validation whenever the measured variable has been used before, satisfying design goal 2.

Secondly, the experimental procedure does not need to be explicitly simulated when unecessary or impractical. This satisfies design goals 2 and 4.

Thirdly, if some aspect of the experiment was omitted in an earlier version of the validation and later deemed relevant to reproducing it in-silico, this approach does not require us to break the existing interface, only to add these new variables for models to take into account or ignore, as deemed necessary by their description of the system. For example, consider a validation for a pattern of synaptic strength, based on an experiment with pair recordings in-vitro. If, after implementing the validation and supporting models, some finding suggests that its results differ from those seen in-vivo due to the concentration of calcium ions in the artificial cerebro-spinal fluid, in a procedural validation we would need to add a \verb+apply_ACSF+ method, breaking compatibility with existing models. With our approach, it is a matter of creating an \verb+EXTRACELLULAR_CALCIUM+ parameter and adding this to the dataframe of observations. This column will simply be ignored by existing models that do not implement its variable, while allowing new models to take advantage of it.

Finally, model methods in sciunit often require fairly complex data as their inputs, the format of which is not well constrained in python. This makes it difficult to guarantee that a method which works for one validation will work for a different one, as the second may use different terminology or formats for the data passed to the method. Devoting effort to the standardization of data rather than specification of procedures prevents this issue from arising in our approach.

In general, the validation requests the high-level feature validated from the model, rather than the lower-level features used to calculate it. For example, if we wanted to validate the firing rate of a population of neurons in a brain region, we would request from the model the firing rates of said neurons. We would not request spike times and calculate firing rates from them, nor would we request local field potentials and calculate spike times from those, as doing so excludes models that do not explicitly represent spike times, for which a firing rate validation may still be relevant (conflicting with design goal 4). In addition, our approach provides greater implementational flexibility to the model, allowing precomputing and caching of firing rates that will be reused across validations.

The obvious downside of this decision is that the transformation from low-level properties to high level ones will need to be implemented anew for each model. We address this by providing, where applicable, general methods for converting from one property to another as a matter of definition, rather than experimental procedure.  These can be used or ignored by a model as suits its implementation. Currently, the implementation of general-purpose methods such as these by users of the framework is optional and poorly incentivised. Improving this is discussed in section \ref{sec:limitations}.

Another design decision we made was to always request all of the measurements required for a validation at once. This way the model has more freedom to make use of caching or commonalities in the requested data to speed up its computations, and it can request all the simulations needed for a validation at once or parallelize its operations. In section \ref{sec:batch-process} we discuss whether the tradeoffs of this approach are worthwhile, or if this decision should be revised in subsequent versions.

In order to enable code reuse between validations and meet design goal 2, each validation is composed of multiple components which can be reused and combined across validations. 

\begin{minipage}{\linewidth}
\begin{lstlisting}[language=Python]
validation_instance = Analysis(
    measured_variable=terms.FIRING_RATE, # can also be list of terms
    observations=pd.read_parquet("path/to/experimental/data.parquet"),
    stats=stats.mann_whitney_u,
    verdict=stats.PooledPvalueThreshold(0.05),
    plotter=plots.hist, # most seaborn-based plotting functionality supported
)
# runs the validation on one or more models
validation_instance(*models)
\end{lstlisting}
\end{minipage}

The plotter and stats objects are both callables of the form:

\begin{minipage}{\linewidth}
\begin{lstlisting}[language=Python]
def plotter(
    data: pd.DataFrame,
    independent: list[Term],
    dependent: list[Term], 
    compare: list[Term]
) -> dict:
    ....
\end{lstlisting}
\end{minipage}

Where the data will be a dataframe of both model and experimental data, independent the column names of the independent variables, dependent the column names of the dependent variable(s) and compare the list of columns defining which datasets to compare (typically, this is [DATASET], containing the experiment and model labels). For the statistical test the resulting dict has the tested null hypotheses as the keys and the dataframe of test statistics as the values. For the plotter, the dict has names of figures as the keys, and instances of \verb+matplotlib.pyplot.Figure+ as the values. The verdict, which determines whether the tested hypotheses pass or fail, recieves the output of the stats function and returns a dict with 'Pass/Fail/NA' for each hypothesis.

For many validations, plotting may be trivially accomplished using the seaborn library \parencite{waskom_seaborn_2021}. To simplify support for this, the plotter may also be a callable of the form \verb+f(data, x, y, hue)->plt.Axes+, so that functions like \verb+seaborn.boxplot+ can be passed as plotters directly without needing a wrapper.

Rather than returning a specific score, a validation returns test statistics and pass/fail states for one or more hypotheses. Typically these hypotheses concern one of two levels of epistemic sufficiency: whether some qualitative trend in the data is reproduced in the model, and whether the data is quantitatively explainable by the model. Validations can be easily modified to test different hypotheses, allowing the evaluation of sufficiency conditions specific to a particular application. 

Alongside these results, experiment and model data is also output for further analysis, as well as figures which help illustrate any mismatches between model and data. It also includes docstrings for the methods used and a list of auxillary assumptions made in the validation process by models and statistical methods (which can be reported by a raising a special subclass of \verb+Warning+). The full validation report can be saved and loaded with the \verb+io+ submodule.

\section{Results}

In this section we will introduce several concrete examples of validating the mouse primary visual cortex model from \cite{dictus_we_2026}, and discuss the challenges encountered in applying this design. The challenges can be grouped into 6 categories.

\begin{enumerate}
    \item In some cases the biases of some experimental methods make it necessary to model those biases or simulate the experimental procedure directly \label{ch:describe-experiment}
    \item The determination of the variable measured by an experiment is theory-laden, and may therefore incorporate assumptions incompatible with some models \label{ch:theory-ladeness}
    \item Practical constraints of a model influence how experimental data should be processed for an optimal comparison, thereby indirectly applying those constraints to models that subsequently use the validation \label{ch:model-constraint}
    \item In some cases the scope of an experiment is not entirely covered by the scope of a model, which can lead to various problems depending on how it is addressed \label{ch:scope-issues}
    \item Some variables, particularly orientation selectivity, can be computed in a variety of different ways in different contexts \label{ch:poorly-defined}
    \item Model-side implementations that can be reused across many validations are often slower than purpose-built implementations, as well as more challenging to write \label{ch:model-inefficient}
\end{enumerate}

Challenges \ref{ch:describe-experiment}, \ref{ch:theory-ladeness}, and \ref{ch:poorly-defined} will be faced by any approach to validation, though operationalism addresses them more explicitly than realism does. Challenge \ref{ch:model-constraint} will be faced by any validation approach, and likely more intensely by an operationalist one, as the constraints of the model used when the validation is first designed with influence what manner of model of the experimental procedure is feasible. Likewise, Challenge \ref{ch:scope-issues} and \ref{ch:model-inefficient} will be faced by any validation method, regardless of its underlying philosophy. In section \ref{sec:limitations} we will discuss how each of these challenges may be addressed in future work.

For each example we will first describe its implementation, and subsequently explain which challenges present themselves. Code snippets are shown for key parts of the model implementation, but most of the code is omitted. The full code can be found at \href{https://github.com/HDictus/pyRMV/tree/rename-module}{here}. 

\subsection{Firing rates}

The first example we consider is of comparing the number of spikes per unit time (firing rate) in models to experiments. For these initial validations we concern ourselves with the 'spontaneous' activity of neurons, in visual neuroscience this is taken to be the activity when the only stimulus presented is a blank gray screen \cite{ma_visual_2010,de_vries_large-scale_2020,siegle_survey_2021}.

We use data here from two datasets, from \cite{siegle_survey_2021} and \cite{ma_visual_2010}. Both use different experimental methods to measure this property. \cite{siegle_survey_2021} uses measurements of extracellular potential to infer the spiking activity of neurons. This method is strongly biased to sample larger neurons and neurons with high firing rates, and sometimes mistakes the activity of multiple neurons for a single neuron. \cite{ma_visual_2010} conversely, uses cell-attached recordings which do not have the same biases. However, it includes only a few types of neurons, has a small sample size, takes place under anesthesia, and admits its method of genetic targeting may have presently unknown biases.

With an operationalist approach, both validations would require different code, describing their experiment and how firing rates were derived from local field potentials and current traces. In contrast, our realist approach allows us to use the same code to predict the firing rate for both experiments, at least in principle. In practice, the biases of the experimental methods make this difficult (challenge \ref{ch:describe-experiment}). As we will argue in \ref{sec:limitations}, the realist approach ultimately allows models to address this limitation with more flexibility than an operationalist framework.

% TODO move to discussion The advantage of our approach is that the model has more freedom in whether it chooses to simulate the experimental procedure, model its sampling biases without simulation (such as probabilistically excluding neurons based on their properties), or take the experiments at face value and accept the limitations of doing so (supporting design goals 2, 3, and 4). The disadvantage is that because the chosen approach is internal to the model rather than the validation it will not be automatically applied to subsequent models, which negatively impacts design goal 1. This issue could be mitigated if users choose to develop reusable modules for experimental bias modeling. Finding ways to support and incentivise this is another goal for future work.

Each validation is defined in three steps:
\begin{enumerate}
    \item standardizing experimental observation data
    \item choosing or implementing a statistical comparison
    \item choosing or implementing a visual comparison (optional)
\end{enumerate}

\subsubsection{Siegle et al. 2021}

We extracted spontaneous firing rates from the raw data. Some of the data is shown in \ref{tab:siegle-spontaneous}.

\begin{table}[!ht]
    \centering
    \caption{first five rows of the table of observations. second table continues with additional columns. bracketed quantities are pandas.Interval objects}
    \label{tab:siegle-spontaneous}
    \begin{tabular}{|p{0.15\linewidth}|p{0.15\linewidth}|p{0.15\linewidth}|p{0.15\linewidth}|p{0.15\linewidth}|p{0.15\linewidth}|}
    \hline
    \hline
        firing rate (Hz) & cell id & layer & region & spiking class & dataset \\ \hline
        0 & 950911195 & ~ & CA1 & RS & Siegle2021 \\ \hline
        0 & 950911266 & ~ & CA1 & FS & Siegle2021 \\ \hline
        0 & 950911286 & ~ & CA1 & RS & Siegle2021 \\ \hline
        0 & 950912928 & ~ & CA1 & RS & Siegle2021 \\ \hline
        0 & 950913031 & L3 & VISam & RS & Siegle2021 \\ \hline
        0 & 950913893 & ~ & grey & RS & Siegle2021 \\ \hline
    \end{tabular}
    %\caption{observation table continued, bracketed quantities are pandas.Interval objects}
    \begin{tabular}{|p{0.15\linewidth}|p{0.15\linewidth}|p{0.15\linewidth}|p{0.15\linewidth}|p{0.15\linewidth}|p{0.15\linewidth}|}
    \hline
    \hline
        visual stimulus  & visual stimulus duration (ms) & visual stimulus stimulus framerate (Hz) & visual stimulus stimulus width (degrees) & visual stimulus stimulus height (degrees) & ~ \\ \hline
        gray & 2.5 & 1000 & (-120, 120] & (-60, 60] & ~ \\ \hline
        gray & 2.5 & 1000 & (-120, 120] & (-60, 60] & ~ \\ \hline
        gray & 2.5 & 1000 & (-120, 120] & (-60, 60] & ~ \\ \hline
        gray & 2.5 & 1000 & (-120, 120] & (-60, 60] & ~ \\ \hline
        gray & 2.5 & 1000 & (-120, 120] & (-60, 60] & ~ \\ \hline
        gray & 2.5 & 1000 & (-120, 120] & (-60, 60] & ~ \\ \hline
    \end{tabular}
\end{table}

\begin{table}[!ht]
    \centering
    \caption{first five rows of the table of experimental variables automatically extracted by the Analysis object.}
    \label{tab:siegle-spont-params}
    \begin{tabular}{|p{0.08\linewidth}|p{0.1\linewidth}|p{0.1\linewidth}|p{0.15\linewidth}|p{0.11\linewidth}|p{0.12\linewidth}|p{0.12\linewidth}|p{0.12\linewidth}|}
    \hline
    \hline
        layer & region & spiking class & spike recording method & visual stimulus  & visual stimulus duration (ms) & visual stimulus stimulus framerate (Hz) & visual stimulus stimulus width (degrees) \\ \hline
        L1 & CA1 & RS & extracellular & gray & 2.5 & 1000 & (-120, 120] \\ \hline
        L1 & VISam & FS & extracellular & gray & 2.5 & 1000 & (-120, 120] \\ \hline
        L2 & VIS & RS & extracellular & gray & 2.5 & 1000 & (-120, 120] \\ \hline
        L2 & VISal & RS & extracellular & gray & 2.5 & 1000 & (-120, 120] \\ \hline
        L2 & VISam & FS & extracellular & gray & 2.5 & 1000 & (-120, 120] \\ \hline
    \end{tabular}
\end{table}

The terminology used in the table is defined in the \verb+.terminology+ module:

\begin{minipage}{\linewidth}
\begin{lstlisting}[language=Python]
#firing rate listed previously

SPIKING_CLASS = Term(
    "spiking class",
    description=(
        "Spiking class of a neuron, either FS or RS"
        "denoting fast-spiking and regular-spiking respectively"
    )
)
REGION = Term(
    "region",
    description=(
        "Acronym of a brain region according to AIBS"
        " atlas naming convention"
    )
)

\end{lstlisting}
\end{minipage}
For the statistical comparison we chose a Mann-Whitney U-test, which tests the null hypothesis that two samples come from the same distribution without making assumptions about the shape or moments of the distributions. If this statistical test passes then the model is considered to accurately represent the target system. The p-value threshold for failure is 0.05, but corrected with a bonferroni correction for the number of tests in the validation.

\begin{minipage}{\linewidth}
\begin{lstlisting}[language=Python]
# in stats.py
def mann_whitney_u(
    data: pd.DataFrame, dependent: str, independent: List[str], compare: str
):
    """Run the Mann-Whitney U-test for the specified data.
# some lines omitted for brevity
    Hypothesis:
        The dependent variable follows the same distribution for both
        compared populations
    """
    hypotheses = {}
    for label1, data1, label2, data2 in _iter_compare(data, compare):
        hypothesis = (
            f"The underlying distribution of {dependent}"
            f" for {label1} and {label2} is the same"
        )
        data2_groups = data2.groupby(independent, dropna=False)

        out_list = []
        for independent_values, grouped1 in data1.groupby(
            independent,
            dropna=False
        ):
            sample1 = grouped1[dependent].dropna().values
            sample2 = grouped2[dependent].dropna().values
            if len(sample1) == 0 or len(sample2) == 0:
                pvalue = np.nan
            else:
                pvalue = stats.mannwhitneyu(
                            sample1,
                            sample2,
                        ).pvalue

            out_list.append({
                **dict(zip(independent, independent_values)),
                terms.PVALUE: pvalue})
        hypotheses[hypothesis] = pd.DataFrame(out_list)
    return hypotheses
\end{lstlisting}
\end{minipage}

This is fairly straightforward, but has several disadvantages. Firstly, it does not distinguish between-subject variation from within-subject variation in the experiment. %A model which presents a plausible instance of a mouse visual cortex may fail the validation because the experimental distribution is broadened by inter-subject variability. Likewise, a model which presents multiple subjects but wrongfully conflates inter-subject with inter-neuron variability can pass the validation. 
Secondly, when many independent validations are run this p-value based approach would lead to even a perfect model failing around 5\% of the validations. We will expand on these limitations and propose solutions in section \ref{sec:limitations}

For the plotting we used a histogram plotter which generates a separate histogram for each combination of independent variables, allowing close inspection of any specific cell type and layer which performs poorly on the validation. 

\begin{minipage}{\linewidth}
\begin{lstlisting}[language=Python]
def hist(data: pd.DataFrame, dependent: str, independent: List[str],
         compare: str) -> Dict[str, plt.Figure]:
    """Create histograms comparing datasets.
    # docstring omitted for brevity
    """
    figs = {}
    independent_vars = independent if len(independent) > 0 else np.zeros(len(data))
    for indvars, alldata in data.groupby(independent_vars):
        fig = plt.figure()
        bins = np.linspace(
            np.nanmin(alldata[dependent]), np.nanmax(alldata[dependent]), 50
        )
        plt.title(str(indvars))
        for label, dataset in alldata.groupby(compare):
            plt.hist(
                dataset[dependent], density=True, label=label, bins=bins, alpha=0.5
            )
        ymax = plt.gca().get_ylim()[1]
        plt.xlabel(dependent)
        plt.legend()
        figs[str(indvars)] = fig
\end{lstlisting}
\end{minipage}

Note that by following the abstracted interface imposed by the framework, these statistical methods are defined without respect to any specific details of the validation.
They can therefore be re-used for any validation with a continuously distributed measurement variable.

Finally, the Analysis is defined by composing these different parts.

\begin{minipage}{\linewidth}
\begin{lstlisting}[language=Python]
    siegle_spontaneous_2021 = Analysis(
        doc="""
        We compare to the firing rate distribution for blank gray stimuli
        observed in Siegle et al. 2021. Note that the experiment used
        extracellular electrodes, and therefore over-estimates firing rates by
        an unknown amount.
        """,
        measurement=terms.FIRING_RATE,
        observations=pd.read_parquet(
            files("analysis_neuro.analyses.data").joinpath("siegle_spont.parquet")
        ),
        plotter=plots.hist,
        stats=stats.mann_whitney_u,
        verdict=stats.PooledPValueThreshold(0.05),
    )
\end{lstlisting}
\end{minipage}

The Analysis class automatically detects the experimental variables, which can be seen in table \ref{tab:siegle-spont-params}.

To support this validation in our model we need to define a \verb+firing_rate+ method which accepts all data on the basis of which the firing rate is to be predicted and returns the appropriate predictions. To this end we need:

\begin{enumerate}
    \item a means of handling each variable our model uses to predict firing rate
    \item a means of running/retrieving an appropriate simulation
    \item a method for calculating firing rates from appropriate simulations
\end{enumerate}

For the first point, this involved creating several filters that select appropriate neurons from the model based on the region and layer variables. Our model had multiple drafts using different conventions for the layers and regions. Early drafts used integers for cortical layers, and later ones strings. An early version used a somatosensory cortex model as a stand-in for visual, and another version appended the hemisphere to the region. Early versions had only a single thalamic region, LGd, while later ones distinguished between LGd-sh and LGd-co. Maintaining compatibility of our validations with older model versions was made comparatively straightforward by this approach because it localized such changes in one part of the code (as opposed to scattering them over all validations). This allowed us to regression test and compare different versions of the model using the same code. For brevity, we only show the filters for the last version.

\begin{minipage}{\linewidth}
\begin{lstlisting}[language=Python]
# in bluebrain_models/cells.py, within CellFilter
...
    def _layer_filter(self, params: dict):
        """Construct layer query."""
        if params[terms.LAYER] == "L23":
            layer = ['2', '3']
        else:
            layer = params[terms.LAYER][1]
        return {"layer": layer} # used to query neurons from the model

    def _region_filter(self, params: dict):
        """Construct region query"""
        region = params[terms.REGION]
        subregions = self._atlas._subregion_acronyms(region)

        return {"region": subregions + [region]}
...

...
\end{lstlisting}
\end{minipage}

Importantly, these filters will be automatically reused for any validation containing region or layer data. This supports design goal 1, as the amount of code needed to support a body of validations will be approximately proportional to the number of variables, which scales sub-linearly with the number of validations as most variables are re-used in many places.

Similarly, variables relating to the stimulus are handled in the process of running, caching, or selecting the simulation.

\begin{minipage}{\linewidth}
\begin{lstlisting}[language=Python]
# simplified for brevity and clarity, not actual code
    def get_simulation(self, params):
        # creates subdirectory for each simulation parameter, in order
        file_path = choose_file_path(self._sims_dir, params)
        if not file_path.exists():
           subprocess.run(_create_sbatch_command(file_path, params)
        return file_path
\end{lstlisting}
\end{minipage}

Finally, we define a method for the firing rates. By requesting firing rates instead of spike times, the validation allows us to pre-compute firing rates before querying cell types to save some time, and additionally allows us to cache firing rates in case they are used in multiple validations.

\begin{minipage}{\linewidth}
\begin{lstlisting}[language=Python]
# simplified, not actual code
    def firing_rate(self, parameters):
        # loads/launches simulations
        sim_paths = self.simulations(parameters)
        rates_per_sim = self.simulations.rates(sim_paths)
        out_rates = []
        for i, params in parameters.iterrows():
            sim_params = {p: v for p, v in params if p in sim_paths}
            sim = sim_paths.etl.q({sim_params}).iloc[0]
            rates = rates_per_sim[sim]
            cells = self.cells(params)
            active_rates = rates.etl.q({
                terms.CELL_ID: cells}
            ))
            missing = pd.DataFrame({
                terms.CELL_ID: cells,
                terms.FIRING_RATE: 0
            }).assign(**params).set_index(terms.CELL_ID)
            missing[terms.FIRING_RATE] = rates_for_cells.groupby(
               terms.CELL_ID)[terms.FIRING_RATE].mean()
            out_rates.append(missing.fillna(0).reset_index())
        return pd.concat(out_rates)
\end{lstlisting}
\end{minipage}

The biases present in extracellular recordings present us with challenge \ref{ch:describe-experiment}. Ideally, the biases of an experimental method are corrected before validation to provide more accurate targets for the model. However, the extent of the biases in these experiments are not known, and determining them is an ongoing area of research \parencite{laquitaine_spike_2024,tharayil_bluerecording_2025,sorrenti_understanding_2021}. Models employed in this research benefit from validating against uncorrected data, not corrected data. Moreover, at least in the case of extracellular recordings, even if the biases were well-understood, it would not be possible to 'correct' the population distribution of firing rates, but at best its mean. The shape of the firing rate distribution affects how information is processed \parencite{buzsaki_log-dynamic_2014}, and is therefore also an important property to validate.

For these reasons it is preferable to account for the biases of these experiments in the process of prediction from the model. We therefore used the experimental data as-is and left whether and how to model its biases to the developer of a particular model. In this way, the details of the experiment provide further specificity on the property measured, and the model decides whether and how to use this specificity.

To quantitatively match this validation, our model would need to mimic the sampling biases of the extracellular electrodes. We did not design a method for this, but the presence of the \verb+terms.RECORDING_METHOD+ variable in the experimental variables would allow future models to detect that extracellular electrodes were used, and consequently apply an approach to model the biases of this method.

\cite{siegle_survey_2021} includes data from brain regions we did not model. Our model will return NaN as the firing rates for these regions, which can lead to unintuitive behavior in pandas dataframes, requiring specific handling in the plotting and statistical analysis methods. This is an instance of challenge \ref{ch:scope-issues}. 

This example also presents us with challenge \ref{ch:model-constraint}. Simulating neural activity for the full duration of the experiment may not be feasible for all models, including many for which spontaneous firing rates are important. For our model, the gray screen duration from \cite{siegle_survey_2021} would be computationally expensive to simulate. While the population average firing rates can be calculated accurately from shorter simulations when there are many neurons, the shape of the firing rate distribution is not preserved in shorter simulations. To understand why, imagine a set of poisson-processes firing at the same rate, and consider how the number of observed events will be distributed within a long or short sampling period.

Since the raw data was available, we addressed this by recomputing spontaneous firing rates using small time windows of 2.5 seconds. This way the validation can be applied to computationally expensive models such as ours without additional assumptions or methods to compensate the distribution. 

\subsubsection{Ma et al. 2010}

Raw data was not available for \cite{ma_visual_2010}, but the population mean firing rates for neurons expressing the genes for Somatostatin and Parvalbumin could be retrieved from their figures. 

The standardized data can be seen in table \ref{tab:ma_spont}.

\begin{table}[!ht]
    \centering
    \caption{Observations table for Ma et al. 2010 validation. Second table continues with additional columns, same rows.  Bracketed quantities are pandas.Interval objects}    
    \label{tab:ma_spont}
    \begin{tabular}{|p{0.2\linewidth}|p{0.2\linewidth}|p{0.2\linewidth}|p{0.2\linewidth}|p{0.2\linewidth}|}
    \hline
    mean firing rate (Hz) & sample size & std firing rate (Hz) & layer & region \\ \hline
    0.082 & 37 & 0.165 & L4 & VISp \\ \hline
    2.889 & 12 & 2.311 & L4 & VISp \\ \hline
    0.165 & 25 & 0.247 & L23 & VISp \\ \hline
    1.651 & 21 & 2.064 & L23 & VISp \\ \hline
    \end{tabular}
    \begin{tabular}{|p{0.2\linewidth}|p{0.2\linewidth}|p{0.2\linewidth}|p{0.2\linewidth}|p{0.2\linewidth}|}
    \hline
    dataset & visual stimulus  & visual stimulus stimulus width (degrees) & visual stimulus stimulus height (degrees) & notes \\ \hline
    Ma2010 & gray & (10, 80] & (-27, 27] & Manually digitized from Figure 1F. \\ \hline
    Ma2010 & gray & (10, 80] & (-27, 27] & Manually digitized from Figure 1F. \\ \hline
    Ma2010 & gray & (10, 80] & (-27, 27] & Manually digitized from Figure 1F. \\ \hline
    Ma2010 & gray & (10, 80] & (-27, 27] & Manually digitized from Figure 1F. \\ \hline
    \end{tabular}
\end{table}

\begin{table}[!ht]
    \centering
    \caption{extracted experimental vars for Ma et al. 2010 validation. Note that additional details can be added to such tables in the future without breaking existing interfaces.}    
    \label{tab:ma_spont_params}
    \begin{tabular}{|p{0.2\linewidth}|p{0.15\linewidth}|p{0.15\linewidth}|p{0.15\linewidth}|p{0.15\linewidth}|p{0.15\linewidth}|}
        \hline
        layer & region & gene expression & visual stimulus  & visual stimulus stimulus width (degrees) & visual stimulus stimulus height (degrees) \\ \hline
        L23 & VISp & Pvalb & gray & (10, 80] & (-27, 27] \\ \hline
        L23 & VISp & Sst & gray & (10, 80] & (-27, 27] \\ \hline
        L4 & VISp & Pvalb & gray & (10, 80] & (-27, 27] \\ \hline
        L4 & VISp & Sst & gray & (10, 80] & (-27, 27] \\ \hline
    \end{tabular}
\end{table}

To use the histogram plot for this example, we just needed to add a few lines:

\begin{minipage}{\linewidth}
\begin{lstlisting}[language=Python]
        if terms.MEAN + dependent in alldata:
            plt.vlines(
                alldata[terms.MEAN + dependent].unique(),
                ymin=0,
                ymax=ymax,
                color="gray",
                linestyle="dashed",
                label="experimental mean",
            )
\end{lstlisting}
\end{minipage}

Since only a mean value was available for the experiment, we used a bootstrapping method to assess the plausability of the firing rate distribution. The \verb+bootstrap_mean+ method could be reused whenever only a population average is reported for the measured variable in an experiment, as can be seen in subsequent sections.

\begin{minipage}{\linewidth}
\begin{lstlisting}[language=Python]
# in stats.py
def bootstrap_mean(
    data: pd.DataFrame,
    dependent: str,
    independent: List[str],
    compare: str,
    num_samples=10000,
):
    """Test whether a mean value could be sampled from a distribution of values.
    # docstring omitted for brevity
    """
    rng = np.random.default_rng(1)
    hypotheses = {}
    for label1, dataset1, label2, dataset2 in _iter_compare(data, compare):
        # some lines emitted for clarity and brevity
        if (
            np.isnan(dataset1[terms.MEAN + dependent]).all() 
            or np.isnan(dataset2[dependent]).all()
        ):
            # not the datasets to compare
            continue

        group_by_independent = dataset2.groupby(independent)
        tests = []
        for group, distr in group_by_independent:
            vals = mean_values.loc[group, [terms.MEAN + dependent, terms.SAMPLE_SIZE]]
            mean, size = vals.values

            samples = rng.choice(
                distr[dependent], size=(int(size), num_samples), replace=True
            )
    
            distrmean = distr[dependent].mean()
            sample_means = samples.mean(axis=0)

            if mean < distrmean:
                pvalue = (sample_means <= mean).mean()
            else:
                pvalue = (sample_means >= mean).mean() 

            tests.append(
                {
                    **dict(zip(independent, group)),
                    terms.PVALUE: pvalue,
                }
            ) 
        hypotheses[
            f"The result of {label1} could be sampled from the same distribution as {label2}"
        ] = pd.DataFrame(tests)
    return hypotheses
\end{lstlisting}
\end{minipage}

The validation is defined as:

\begin{minipage}{\linewidth}
\begin{lstlisting}[language=Python]
ma_spontaneous_2010 = Analysis(
    observations=importlib.import_module(
        "analysis_neuro.analyses.data.ma_2010"
    ).spontaneous,
    measurement=terms.FIRING_RATE,
    plotter=plots.hist,
    stats=stats.bootstrap_mean,
    verdict=stats.PooledPValueThreshold(0.05),
)
\end{lstlisting}
\end{minipage}

Due to the similarities with the previous validation, supporting this validation for our model only required us to implement support for the gene expression variable. Our model had a gene class associated with each neuron. A model without gene-based classification could instead introduce additional assumptions to predict the properties of a genetic class from morpho-electrical types which frequently express the gene.

\begin{minipage}{\linewidth}
\begin{lstlisting}[language=Python]
    def _gene_expression_filter(self, params: dict):
        """Get cells expressing a particular gene based on their mclass."""

        mclasses = self.mclasses
        ge = params[terms.GENE_EXPRESSION]
        if ge not in ['Sst', 'Pvalb', 'Vip', 'Lamp5']:
            return {'node_id': []} # will give empty result
        return {'mclass': [mc for mc in mclasses if ge in mc]}
\end{lstlisting}
\end{minipage}

All other model-side code was already created for the previous validation, and can be readily reused. As a model increases the number of different variables it supports it becomes increasingly likely that any new validation is supported without any code edits being necessary at all. 

If we wish to simulate the effects of anesthesia in our model, we would need to add the relevant anesthesia variables to the validation. Any other models already using the validation would continue to work with it, as no new functionality is required of them. If we had used an operationalist approach, in contrast, we would then need all models supporting the validation to implement a \verb+apply_anaesthetic+ method in order to run the validation, breaking compatibility. In contrast, our realist model-validation interaction allows the interface to remain more stable if additional details are considered relevant.

\subsection{Orientation selectivity}

Some neurons in the primary visual cortex respond preferentially to stimuli of specific orientations. This is called orientation selectivity, and it has been studied extensively and is the basis for a lot of theoretical work. Consequently we included a validation of the extent to which orientation selectivity is present in a model. We used experimental data from \cite{siegle_survey_2021} for this.

In this case the measured property is orientation selectivity. This is typically quantified with an orientation selectivity index (OSI). Different OSIs exist, and different stimuli can be used to determine them. Similarly, they can be applied to different measures of neural activity, such as membrane potential, and various measures of firing rate. Some uses of the orientation selectivity index first subtract a baseline, such as spontaneous activity, from the response variable. Orientation selectivity validations (and similar validations such as direction selectivity) therefore present us with challenge \ref{ch:poorly-defined}.

While firing rate responses had been quantified in response to a specific stimlus, an OSI is necessarily in response to a set of stimuli. Multiple stimuli could not be readily represented by primitive data types in the dataframe. To solve this we created the \verb+dataframe_pointer+ package, which allowed us to store an immutable, hashable reference to a dataframe of multiple stimuli in a STIMULI column.

The data could then be standardized and loaded with:

\begin{minipage}{\linewidth}
\begin{lstlisting}[language=Python]
osi = pd.read_parquet(file(<filepath>))
osi[terms.STIMULI] = pd.DataFrame({
   terms.VISUAL_STIMULUS + terms.STIM_ORIENTATION: np.arange(0., 360., 45.),
   terms.VISUAL_STIMULUS + terms.ANGLE_AZIMUTH: pd.Interval(-120, 120),
   ....
}).pointer()
\end{lstlisting}
\end{minipage}

The validation could be defined as follows:

\begin{minipage}{\linewidth}
\begin{lstlisting}[language=Python]
siegle_osi_tf4_2021 = Analysis(
    doc="""
    We compare to the levels of orientation selectivity observed in
    Seigle et al. 2021 for drifting gratings with a temporal
    frequency of 4 Hz
    """,
    measurement=terms.ORIENTATION_SELECTIVITY,
    observations=importlib.import_module(
        "analysis_neuro.analyses.data.siegle_2021"
    ).osi,
    plotter=sns.boxplot,
    stats=stats.mann_whitney_u,
    verdict=stats.PooledPValueThreshold(0.05),
)
\end{lstlisting}
\end{minipage}

\begin{table}[!ht]
    \centering
    \caption{Five rows of the observations table for the orientation selectivity validation}
    \begin{tabular}{|p{0.25\linewidth}|p{0.25\linewidth}|p{0.25\linewidth}|p{0.25\linewidth}|}
    \hline
        gid & orientation selectivity index & layer & spiking class \\ \hline
        950907362 & 0.423 & L5 & FS \\ \hline
        950907364 & 0.243 & L5 & RS \\ \hline
        950907471 & 0.455 & L3 & RS \\ \hline
        950907524 & 0.128 & ~ & FS \\ \hline
        950907526 & 0.401 & ~ & FS \\ \hline
    \end{tabular}
    \begin{tabular}{|p{0.25\linewidth}|p{0.25\linewidth}|p{0.25\linewidth}|p{0.25\linewidth}|}
    \hline
        region & dataset & spike recording method & stimuli \\ \hline
        VISam & Siegle2021 & extracellular & dataframe pointer:    [8 rows x 7 columns] \\ \hline
        VISam & Siegle2021 & extracellular & dataframe pointer:    [8 rows x 7 columns] \\ \hline
        VISam & Siegle2021 & extracellular & dataframe pointer:    [8 rows x 7 columns] \\ \hline
        MB & Siegle2021 & extracellular & dataframe pointer:    [8 rows x 7 columns] \\ \hline
        MB & Siegle2021 & extracellular & dataframe pointer:    [8 rows x 7 columns] \\ \hline
    \end{tabular}
\end{table}

\begin{table}[!ht]
    \centering
    \caption{first five lines of parameters table}
    \begin{tabular}{|p{0.2\linewidth}|p{0.2\linewidth}|p{0.2\linewidth}|p{0.2\linewidth}|p{0.2\linewidth}|}
    \hline
        layer & spiking class & region & spike recording method & stimuli \\ \hline
        L1 & FS & VISam & extracellular & dataframe pointer:    [8 rows x 7 columns] \\ \hline
        ~ & RS & CA1 & extracellular & dataframe pointer:    [8 rows x 7 columns] \\ \hline
        L2 & FS & VISam & extracellular & dataframe pointer:    [8 rows x 7 columns] \\ \hline
        L2 & RS & VIS & extracellular & dataframe pointer:    [8 rows x 7 columns] \\ \hline
        L2 & RS & VISal & extracellular & dataframe pointer:    [8 rows x 7 columns] \\ \hline
    \end{tabular}
\end{table}

In most studies a baseline firing rate, such as the spontaneous firing rate, is subtracted from the responses before the OSI is calculated. If the response of a neuron to a stimulus is less than its baseline value, this will lead to a negative number. Additionally, this requires spontaneous activity simulations whenever orientation selectivity is to be assessed. Since the raw neural responses were included in the experiment we could recompute the OSI using a different method. Instead of subtracting spontaneous activity, we subtracted the smallest firing rate observed in the stimulus set, preventing negative values and allowing OSI to be calculated from grating responses alone.

We implemented a general-purpose method for measuring this orientation selectivity index on any model which represents firing rates in response to drifting gratings. However, for comparisons to experiments where the raw data are not available it may not be reasonable to reuse this method, as it may calculate OSI in a slightly different way.

\begin{minipage}{\linewidth}
\begin{lstlisting}[language=Python]
# in measurements/orientation_selectivity.py

def from_firing_rate(model, parameters):
    """Measure orientation selectivity on the basis of some response property (e.g. Firing rate).
    # docstring omitted for brevity
    """
    out = []
    for _, row in tqdm(parameters.iterrows(), total=len(parameters)):
        stimuli_shown = row[terms.STIMULUS].df
        response = model.firing_rate(stimuli_shown.assign(**row)).reset_index()
        minfr = response.groupby(terms.CELL_ID)[terms.FIRING_RATE].min()
        response[terms.FIRING_RATE] -= minfr.loc[response[terms.CELL_ID]].values        
        if len(response) == 0 or np.all(np.isnan(response[response_measurement])):
            continue
        selectivity = g_OSI_signal(
            response[terms.FIRING_RATE],
            response[terms.VISUAL_STIMULUS + terms.STIM_ORIENTATION],
            groupby=terms.CELL_ID
        )
        selectivity.name = terms.ORIENTATION_SELECTIVITY
        out.append(selectivity.reset_index().assign(**row))

    return pd.concat(out, axis=0)
\end{lstlisting}
\end{minipage}

In order to verify whether the level of orientation selectivity in the model was attributable to chance, we wished to compare the model to a control model in which the firing rates of different simulations were shuffled. Due to the validation not having direct access to the firing rates, it was necessary to perform this shuffling within the model class and pass it as a separate model, so that the statistical comparison would also be applied to it and the full model. An implementation which accesses firing rates could have implemented this as a part of the statistical analysis.

Challenge \ref{ch:model-constraint} was posed even more strongly in this case than in the preceding example. In addition to presenting stimuli for many trials over long durations, \cite{siegle_survey_2021} presented drifting grating stimuli at many different temporal frequencies. To prevent the simulation time from becoming too large we recomputed the OSI using only stimuli with a temporal frequency of 4Hz. We selected this frequency because simulation results from \cite{billeh_systematic_2020} at this frequency were also available, allowing us to compare it to our model. For experiments where the raw data are not available, it would instead be necessary to compare the values reported in the experiments to those computed from limited simulations in the model, leading to a less meaningful validation. 

Challenge \ref{ch:model-inefficient} also presented itself in this validation. Initially we needed to load simulations anew for each cell type and compute the firing rates from that. A standalone implementation of this validation could simply load all the simulations, calculate the orientation selectivity indices, and finally divide the cells into types. However, the generalizable construction of our validations prevented this. Much of the computational costs could be offset by caching firing rates for each simulation, but the overall implementation was more complex as a result.

One of the mechanisms underlying orientation selectivity is within the pattern of input to cortical neurons from the thalamus. \cite{lien_tuned_2013} quantified this by showing drifting gratings to a mouse while silencing the cortex and recording currents in cortical neurons. They found that the degree to which the input current was modulated by the frequency of the stimulus was dependent on orientation, and quantified this tendency with the orientation selectivity index of the amplitude of a sine wave fit to the input current (F1-OSI), averaging 0.23. Conversely, they observed very little orientation selectivity of the mean current, averaging 0.027.

Reproducing these degrees of orientation selectivity in the input signal constrains a model to reproduce orientation selectivity in a realistic manner, and validating them is therefore valuable. The OSI in this case is of a different signal, and calculated in a different manner. We obviously cannot use the existing orientation selectivity code for it.

% TODO: this actually isn't correct - if we define different orientation selectvity indices, we can reuse the definitions
% so we don't have this weakness compared to an operationalist approach
The variation in the different ways that orientation selecitvity can be computed poses challenge \ref{ch:poorly-defined}, which an operationalist approach would partly mitigate. With an operationalist approach the procedure for calculating an experiment's particular version of OSI is contained in the validation. With our realist approach, treating the OSI as a real property of the target system, we cannot do this, and must instead define multiple different orientation selectivity indices, and handle them separately within each model. In \ref{sec:limitations} we discuss how this should be addressed as the library of validations expands.

The validation for F1-OSI was implemented by defining the term \verb+OSI_CURRENT_FM+, and subsequently defined as follows:

\begin{minipage}{\linewidth}
\begin{lstlisting}[language=Python]
lien_osi_fm_2013 = Analysis(
    measurement=terms.OSI_CURRENT_FM,
    observations=pd.DataFrame({
        terms.MEAN + terms.OSI_CURRENT_FM: [0.23],
        terms.SAMPLE_SIZE: 13,
        terms.POSTSYNAPTIC + terms.REGION: "VISp",
        terms.SPECIES: "mouse",
        terms.VOLTAGE_CLAMP: -70,
        terms.POSTSYNAPTIC + terms.LAYER: "L4",
        terms.PRESYNAPTIC + terms.REGION: "LGd",
        terms.DATASET: 'Lien2013',
        terms.MTYPE: "PC",
        terms.STIMULI: importlib.import_module(
            "analysis_neuro.analyses.data.lien_2013"
        ).stimulus.pointer(),
        terms.RESPONSE_CLASS: "sON/tOFF",
    }),
    plotter=plots.hist,
    stats=stats.bootstrap_mean,
    verdict=stats.PooledPValueThreshold(0.05)
)
\end{lstlisting}
\end{minipage}

We also created a function for calculating the amplitude of the frequency modulation, which models can apply or reimplement as preferred.

\begin{minipage}{\linewidth}
\begin{lstlisting}[language=Python]
from scipy import optimize
import numpy as np

def frequency_modulation_amplitude(time, signal, frequency):
    """Amplitude of frequency modulation at the specified frequency"""
    signal = signal - signal.mean()
    def sin_at_freq(time, offset, amplitude):
        return np.sin((time * 2 * np.pi * frequency) + offset) * amplitude
    (offst, amp), _ = optimize.curve_fit(sin_at_freq, time, signal)
    return np.abs(amp)
\end{lstlisting}
\end{minipage}
% TODO see exactly here, an example of how we can still get the benefits of operationalism like this
Our simulator did not permit the measurement of currents from specific sources during simulation, and so we would be required to re-run simulations with other sources of excitation turned off to record current from LGd. Additionally, we wished to apply this validation to the pattern of thalamocortical connectivity before the cell models and synaptic parameters had been finalized. The flexibility of the realist approach permitted us to use a histogram of incoming spiking activity as a stand-in for current, solving both issues.

\subsection{Connection probability}

Alongside validating the response properties of our model, we wished to validate its connectivity as well. Our connectivity was constructed based on a number of assumptions, and checking that these assumptions reproduce connectivity trends allows us to assess the adequacy of those assumptions. Additionally, validating the connectivity of the model grants us greater certainty that the mechanisms by which it predicts responses resemble those of the target system.

Connection probability, unlike the activity variables, does not describe the value of a concrete instance of a property. That is to say, connection probability is not defined for any particular pair of neurons: they are either connected or they are not. Connection probability is only definable for a non-specific pair or population of pairs.

Connection probability can be measured in a variety of ways; and an operationalist framework would require novel code for each of these. Because our realist framework allows us to omit details of the experiment not relevant to our model, we could use th same code for different experiments.

One connection-probability validation we implemented used the data from \cite{schneider-mizell_inhibitory_2025}. This dataset includes connection probabilities to and from inhibitory neurons in a columnar subset of the mouse visual cortex. The diameter of this column (100um) is an important variable in the validation, as the connection probability observed in a column will decrease with its size because sampled neurons are more distant from one another. This dataset used its own, novel classification system for neurons, neccessitating the creation of a new term to represent it. Some of the standardized data can be seen in Table \ref{tab:smizell-connprob}.

\begin{table}[!ht]
    \centering
    \caption{Five rows of the observations table for validating connection probability from \cite{schneider-mizell_inhibitory_2025}}
    \label{tab:smizell-connprob}
    \begin{tabular}{|p{0.25\linewidth}|p{0.25\linewidth}|p{0.25\linewidth}|p{0.25\linewidth}|}
    \hline
    \hline
        Presynaptic layer & Postsynaptic layer & Presynaptic Schneider-Mizell type & Postsynaptic Schneider-Mizell type \\ \hline
        L1 & L1 & STC & STC  \\ \hline
        L1 & L2 & STC & DTC \\ \hline
        L1 & L2 & STC & L2a \\ \hline
        L1 & L2 & STC & L2b \\ \hline
        L1 & L2 & STC & L2c \\ \hline
    \end{tabular}
    \begin{tabular}{|p{0.25\linewidth}|p{0.25\linewidth}|p{0.25\linewidth}|p{0.25\linewidth}|}
    \hline
    \hline
        sample size & column radius (um) & connection probability & dataset \\ \hline
        72 & 50 & 0.333 & Schneider-Mizell2025 \\ \hline
        27 & 50 & 0.111 & Schneider-Mizell2025 \\ \hline
        513 & 50 & 0.081 & Schneider-Mizell2025 \\ \hline
        594 & 50 & 0.062 & Schneider-Mizell2025 \\ \hline
        648 & 50 & 0.072 & Schneider-Mizell2025 \\ \hline
    \end{tabular}
\end{table}

We chose a binomial test, which would assume that the model provides its "true" estimate of connection probability, and subsequently assesses the probability of sampling as many connected pairs as occurred in the experiment. As before, a bonferroni-corrected p-value above 0.05 is considered a pass.

%Because connection probability is a statistic over multiple pairs of neurons, a model may choose to represent connection probability not with instances of connections, but with a continuous distance-dependent function (especially if the model is operating at the level of neuron populations rather than individual neurons). Our way of defining connection probability validations does not smoothly map onto this class of model. Rather, it would be easier in that case to evaluate the probability function at the distances observed in the data and thereby evaluate its likelihood. It remains feasible to translate such models and compare them to representative models, which is not the case with an operationalist approach.

We implemented a reusable method for calculating connection probability that can be applied to any model that explicitly represents neurons and connections between them. We defined an intermediate measurement variable, pair weight, and defined connection probability in terms of it:

\begin{minipage}{\linewidth}
\begin{lstlisting}[language=Python]
# in measurements/connection_probability.py
def from_pair_weights(model, parameters):
    """Measure terms.CONNECTION_PROBABILITY from pair weights.
    # docstring omitted for brevity
    """
    edges = model.pair_weight(parameters)
    edges['conn'] = edges[terms.PAIR_WEIGHT] > 0
    groups = edges.groupby(list(parameters.columns))['conn']
    connprob = pd.DataFrame({
        terms.CONNECTION_PROBABILITY: groups.mean(),
        terms.SAMPLE_SIZE: groups.count()
    }).reset_index()
    return connprob
\end{lstlisting}
\end{minipage}

While this simplifies the process of calculating predicted connection probabilities for other models, the bulk of the complex logic is in the process of retrieving pairs of cells and their weights. We explore in section \ref{sec:limitations} how some changes in the framework could permit code reuse in this part as well.

For other connectivity validations we implemented we similarly defined connectivity statistics in terms of either pair weight or connection weight, allowing models to validate with respect to many different connectivity properties once they implement cell and pair filtering. In order to apply this validation to our model we needed to map their neuron classes onto ours. Applying their classification procedure would have taken a lot of additional work, while the flexibility of our framework allowed us to use a simpler solution.

\begin{minipage}{\linewidth}
\begin{lstlisting}[language=Python]
   def _smizell_type_filter(self, params: dict):
        smizell_type = params[terms.SMIZELL_TYPE]
        inh_mclasses = {
            'DTC': 'Sst',
            'PTC': 'Pvalb', 
            'ITC': 'Vip',
            'STC': 'Lamp5'
        }
        try:
            mclass = inh_mclasses[smizell_type]
            mclasses = concat_populations(
                self._circuit.nodes.get, properties=['mclass']
            )['mclass'].unique()
            return {'mclass': [mc for mc in mclasses if mclass in mc]}
        except KeyError:
            return {'mtype': smtype_to_mtype[smizell_type]}
\end{lstlisting}
\end{minipage}

For limiting cells to a column, we defined a new cell filter:

\begin{minipage}{\linewidth}
\begin{lstlisting}[language=Python]
# in cells.py
    def _column_radius_filter(self, param: dict):
        """Filter cells in a column of a given radius, centered on a region.

        requires both REGION and COLUMN_RADIUS from analysis_neuro.terminology
        to be in param.
        """
        if self._check_cache(param):
            return self._check_cache(param)

        displacement_from_center = self._displacement_from_center(param)
        perpendicular_component_size = np.linalg.norm(displacement_from_center, axis=-1)
        in_column = perpendicular_component_size < param[terms.COLUMN_RADIUS]
        ids = displacement_from_center[in_column].index
        query = {'node_id': list(ids.get_level_values(1))}
        self._columns[column_id] = query
        return query
        ...
     # passing column_num in pair queries lets us retrieve pairs from
     # multiple columns, increasing our sample size
     def _position_to_displacement(self, param):
         if 'column_num' in param:
            center = self._choose_random_center(param)
        else:
            center = positions.mean(axis=0)
        return self._position_to_displacement(positions, center)
\end{lstlisting}
\end{minipage}

%However, when filtering of pairs of neurons is not required for a validation, such as when connection probability is measured within a column, it is far more efficient to calculate connection probability from the number of potential and actual connections than to explicitly represent the pairs of neurons. \todo{so what?}
In predicting connection probability we chose to sample multiple columns of the provided radius and pool their results into a single connection probability estimate per pathway. It would also have been valid to report a separate measurement for each column. Since comparing the output of these two different decisions requires different statistical methods, the ambiguity allowed by our framework poses a risk to the reusability of model methods across validations. An operationalist approach could centralize any such decision in the validation code, thereby circumventing this problem.

Our model had to go pathway-by-pathway, identify presynaptic neurons, identify postsynaptic neurons, apply any distance-based filtering to the pairs.

\begin{minipage}{\linewidth}
\begin{lstlisting}[language=Python]
# for illustrative purposes, not actual code
    for pre, post in iter_pathways(exp_vars):
        pre_cells, post_cells = get_cells(pre), get_cells(post)
        pairs = filter_pairs(np.prod(pre_cells, post_cells))
        pairs['weight'] = _get_weight(pairs).fillna(0)
\end{lstlisting}
\end{minipage}

Not only do the same cell types need to be repeatedly filtered, but each pair of neurons must be filtered anew for each different distance bin. While the costs of this could be greatly reduced by caching, the result was still slower than if we had grouped pairs by the relevant variables instead:

\begin{minipage}{\linewidth}
\begin{lstlisting}[language=Python]
# for illustrative purposes, not actual code
neurons = circuit.get(subset, properties=['mclass', 'flat_x', 'flat_y'])
positions = neurons[['flat_x', 'flat_y']]
distances = np.linalg.norm(
    positions.values - positions.values[np.newaxis, ...],
    axis=-1
)
pairs = pd.Dataframe(
   {'distance_bin': pd.cut(distances.flatten(), range(0, 300, 25))},
    index=pd.MultiIndex.from_product((all_neurons.index, all_neurons.index)),
)
pairs['connected'] = is_connected(pairs.index)
pairs['pre_mclass'] = neurons.lofc[pairs.index.get_level_values(0), 'mclass']
pairs['post_mclass'] = neurons.loc[pairs.index.get_level_values(1), 'mclass']
cols = ['pre_mclass', 'post_mclass','distance_bin']
connprob_data = pairs.groupby(cols)['connected'].mean()
\end{lstlisting}
\end{minipage}

The reasons this approach cannot be applied in a method that is to be reused across multiple validations  are
\begin{enumerate}
    \item the validation may only refer to a small subset of neural types
    \item the classification scheme will vary from experiment to experiment
    \item within the same experiment, classifications may overlap
    \item we may wish to filter by other properties than distance
\end{enumerate}

The purpose-built implementation is fast both to write and to run because it exploits a-priori knowledge about the scope of the computation. Because we choose to request all measurements in a validation at once, it is in theory possible to get similar vectorization benefits in a reusable implementation, but the design of such an implementation would be substantially more complex.

Despite its weaknesses, the row-by-row approach is more expressive and memory-efficient than the batch-processing approach, a point we will return to in section \ref{sec:limitations}. 

This example also presented us with challenge \ref{ch:scope-issues}. Our mapping of excitatory Schneider-Mizell types to the morphological types in our model is a very rough approximation. As a result, a poor match with the experiment may be related to inaccuracies in this mapping rather than in the overall connectivity in the model. It may therefore have been preferable to treat all excitatory neurons within a layer as a single type, without differentiating subtypes. While reconfiguring the validation to work in this way is straightforward (simply adjusting the observations dataframe is sufficient), the goal of these validations is that they can be applied without modification to any relevant model.

One way to resolve this is to regard the two different validations as validating two different things - overall excitatory connection probability and subtype-specific connection probability. Involving both validations in the framework is therefore acceptable, and models which do not contain subtype specificity will simply perform poorly on that validation.

\subsection{Fraction excitation per connection}

One aspect of the connectivity in the model we wished to evaluate is how many neurons in the dorsal lateral geniculate nucleus (LGd) converge on each neuron in the primary visual cortex. For this we compared to \cite{lien_cortical_2018}. An operationalist approach would require us to simulate their procedure - blocking local activity in the cortex, attaching recording pipettes, stimulating individual neurons in the LGd, etc. The realist approach permits a much simpler validation - we can simply look at how many neurons converge in the connectivity of the model, and report this as the model's prediction. However, this simplicity hides a difficulty presented by the interperetation of the experiment.

Thus far we have been committed to validating the model with respect to the property the experiment aimed to infer. We departed from this when validating the number of thalamocortical connections to each cortical neuron. This departure is instructive of the limitations of this principle.

\cite{lien_cortical_2018} measured the amount of excitation contributed by single thalamo-cortical connections and compared this to the total thalamo-cortical current. They divided the mean of single-connection current by the mean of total current and thereby calculated that each cortical neuron receives input from around 80 thalamic afferents. However as pointed out by \cite{ringach_sparse_2021}, if the distribution of single currents is long-tailed (as it is in our model), then it is possible for the true number of afferents to be much smaller. 

\begin{minipage}{\linewidth}
\begin{lstlisting}[language=Python]
lien_fraction_excitation_2018 = Analysis(
    observations=pd.DataFrame({
        terms.POSTSYNAPTIC + terms.REGION: "VISp",
        terms.POSTSYNAPTIC + terms.SYNAPSE_CLASS: "EXC",
        terms.PRESYNAPTIC + terms.REGION: "LGd",
        terms.MEAN + terms.FRACTION_EXCITATION_PER_CONNECTION: [0.012],
        terms.SAMPLE_SIZE: 14,
        terms.DATASET: "Lien2018",
    }),
    measurement=terms.FRACTION_EXCITATION_PER_CONNECTION,
    stats=stats.bootstrap_mean,
    plotter=plots.hist,
    verdict=stats.PooledPValueThreshold(0.05),
)
\end{lstlisting}
\end{minipage}

\begin{minipage}{\linewidth}
\begin{lstlisting}[language=Python]
# in measurements/fraction_excitation.py
def from_connection_weights(model, parameters):
    """Measure fraction excitation per connection from connection weights.
    # omitted for brevity
    """

    edges = model.connection_weight(parameters)
    edges = edges[edges[terms.CONNECTION_WEIGHT] != 0]
    groupcols = list(parameters.columns) + [terms.POSTSYNAPTIC + terms.CELL_ID]
    tot_exc = edges.groupby(groupcols, dropna=False)[
        terms.CONNECTION_WEIGHT
    ].sum()
    idx_cols = groupcols + [terms.PRESYNAPTIC + terms.CELL_ID]
    fin = edges.set_index(idx_cols)[terms.CONNECTION_WEIGHT] / tot_exc
    fin.name = terms.FRACTION_EXCITATION_PER_CONNECTION
    return fin.reset_index()
\end{lstlisting}
\end{minipage}

In this case what the experimenters aimed to measure was convergence, the number of presynaptic neurons converging on each postsynaptic neuron. However, due to Ringach's criticism of the underlying assumptions we chose to validate with respect to fraction of excitation per connection instead.

%(I think I need some sort of formal description of the inverse problem presented by the model and validation)

This case cleary illustrates challenge \ref{ch:theory-ladeness}. Choosing the measured variable for a validation inherently theory-laden, as is the choice to accept an experimenter's framing of the experiment: it depends on a number of other beliefs about the target system. This presents a major challenge to systematic validation in general: as different models implement different beliefs about the target system and so may be logically incompatible with the inferred results of an experiment conducted under different assumptions. This issue is more easily circumvented with strict operationalist approaches, but only if their description of the experiment is exhaustive. We discuss how this should be addressed in section \ref{sec:limitations}.

\subsection{Relative innervation strength}

\cite{ji_thalamocortical_2016} provides data on the strength of input from the dorsal lateral geniculate nucleus (LGd) to different cell types in the primary visual cortex. The experiment proceeded by slicing the brain so that a section of thalamus and the connected section of visual cortex are both present, stimulating the whole thalamus in a short pulse, and recording the observed currents in the visual cortex.

\begin{table}[!ht]
    \centering
    \caption{first five rows of the table of observations.}
    \label{tab:ji-relative}
    \begin{tabular}{|p{0.15\linewidth}|p{0.15\linewidth}|p{0.15\linewidth}|p{0.15\linewidth}|p{0.15\linewidth}|p{0.15\linewidth}|}
    \hline
    \hline
        Presynaptic region & Postsynaptic layer & Postsynaptic region & Postsynaptic gene expression & Postsynaptic synapse class \\ \hline
        LGd & L1 & VISp & ~ & ~ \\ \hline
        LGd & L1 & VISp & ~ & ~ \\ \hline
        LGd & L1 & VISp & ~ & ~ \\ \hline
        LGd & L1 & VISp & ~ & ~ \\ \hline
        LGd & L23 & VISp & PV & ~ \\ \hline
    \end{tabular}
    %\caption{observation table continued, bracketed quantities are pandas.Interval objects}
    \begin{tabular}{|p{0.15\linewidth}|p{0.15\linewidth}|p{0.15\linewidth}|p{0.15\linewidth}|p{0.15\linewidth}|p{0.15\linewidth}|}
    \hline
    \hline
        relative to Presynaptic region & relative to Postsynaptic region & relative to Postsynaptic layer & relative to Postsynaptic synapse class & dataset \\ \hline
        LGd & VISp & L4 & EXC & Ji2016 \\ \hline
        LGd & VISp & L4 & EXC & Ji2016 \\ \hline
        LGd & VISp & L4 & EXC & Ji2016 \\ \hline
        LGd & VISp & L4 & EXC & Ji2016 \\ \hline
        LGd & VISp & L4 & EXC & Ji2016 \\ \hline
    \end{tabular}
    \begin{tabular}{|p{0.15\linewidth}|p{0.15\linewidth}|p{0.15\linewidth}|p{0.15\linewidth}|p{0.15\linewidth}|p{0.15\linewidth}|}
    \hline
    \hline
        connection weight type & extracellular calcium (mM) & ~ & ~ & ~ \\ \hline
        uPSC & 2 & ~ & ~ & ~ \\ \hline
        uPSC & 2 & ~ & ~ & ~ \\ \hline
        uPSC & 2 & ~ & ~ & ~ \\ \hline
        uPSC & 2 & ~ & ~ & ~ \\ \hline
        ~ & ~ & ~ & ~ & ~ \\ \hline
    \end{tabular}
\end{table}

To predict this quantity, a model could proceed to simulate this full procedure, limiting its simulation to a slice, making all LGd neurons within the slice spike once, and recording currents in cortical neurons. However, for some models it is also possible to predict this without simulation by analyzing the parameters of the connectivity. Additionally, if we consider the experiment to be measuring the relative strength of input to different neuron types, using the reasonable assumption that all cell types are equally affected by the slicing procedure, we can remove the need to account for the effect of slicing altogether, as \cite{billeh_systematic_2020} did. Our realist approach permits these optimizations to be made, while an operationalist description of the experiment will make this either more complicated or outright impossible for a model to do. However, using the connectivity in this way involves correctness assumptions that are not necessary when explicit simulation is used (challenge \ref{ch:poorly-defined}), as well as embedding more assumptions about the target system in the validation (challenge \ref{ch:theory-ladeness}).

Because we could interperet the experiment this way, we could support the validation with a reusable connectivity-based method:

\begin{minipage}{\linewidth}
\begin{lstlisting}[language=Python]
def from_connection_weights(model, parameters):
    """Measure relative excitation from connection weights.
    # omitted for brevity
    """
    cond_per_tgid = _weights_sum(model, parameters).reset_index()
    relativecols = [
        col for col in cond_per_tgid 
        if col.startswith(terms.RELATIVE_TO)
        ]
    normalized = []
    for grp, conds in cond_per_tgid.groupby(relativecols, dropna=False):
        relative_params = {
            col.replace(terms.RELATIVE_TO, ''): val
            for col, val in zip(relativecols, grp)
        }
        relative_to = _weights_sum(
            model,
            pd.DataFrame(relative_params, index=[0])
        )
        conds[terms.RELATIVE_EXCITATION] = (
            conds[terms.CONNECTION_WEIGHT] / 
            relative_to.mean()
        )
        normalized.append(conds.drop(columns=[terms.CONNECTION_WEIGHT]))
    return pd.concat(normalized)

def _weights_sum(model, parameters):
    """Calculate sum of connection weights per postsynaptic cell.
    # omitted
    """

    edges = model.connection_weight(parameters)
    groupcols = list(parameters.columns) + [terms.POSTSYNAPTIC + terms.CELL_ID]
    cond_per_tgid = edges.groupby(groupcols, dropna=False)[
        terms.CONNECTION_WEIGHT
    ].sum()
    return cond_per_tgid
\end{lstlisting}
\end{minipage}

%\todo{there are additional points to be made about potential code reuse with \verb+RELATIVE_TO+ and converting back and forth between \verb+CONNECTION_WEIGHT+ and \verb+PAIR_WEIGHT+}

The interperetation of this experiment as measuring relative synaptic strength depends on the assumption that synaptic release probabilities are similar in synapses targeting all the cell types measured. The model from \cite{billeh_systematic_2020} did not model short-term plasticity, and it was therefore natural to accept it. However, synapses from LGd differ in the degree of short-term plasticity \parencite{kloc_target-specific_2014}, and therefore most likely in their release probability. What is more, the concentration of calcium in the artificial cerebro-spinal fluid (2mM) strongly influences release probability, and this differs from in-vivo \parencite{borst_low_2010}. The strength of the effect of calcium concentration varies depending on cell type \parencite{markram_reconstruction_2015}. 

We addressed this by including columns for the type of synaptic weight measured and the extracellular calcium. This allowed the model to multiply connection weights by release probability calculated at that calcium concentration. However, this requires some additional branching logic within the model class which will ultimately calculate some measure of unitary PSC amplitude and pass it forward to the method for measuring synaptic strength. Had we interpreted the validation as giving relative afferent PSC amplitude the model-side code would be simpler and more expressive, and the relationship of the model to the target system more transparent.

\section{Discussion}

The examples we have provided demonstrate both strengths and limitations of the pyRMV framework. The problems a general-purpose validation framework must address are the same regardless of its approach; the approach is simply the means by which those problems are addressed. The core issues that need to be addressed are code/effort duplication between different models (different implementations, different scopes, different conceptual underpinnings) and code/effort duplication between different validations. In addressing this, the solution must have a means of dealing with holism and theory-ladeness, and be as simple to use as possible. Each approach will make different compromises between the core functionality and these other concerns.

%Naively, an operationalist approach can address this by describing the operations involved in an experiment rather than referring to the variables the experimenter interpreted these operations to measure. However, for many experiments it is impractical to exhaustively describe the experiment, requiring many parts of the procedure to be simplified or abstracted, in the process, thereby reintroducing theory-ladeness.

\subsection{Framework strengths}

The clearest strengths of pyRMV lie in how thoroughly it addresses code duplication. The validation-independent form of the statistical and plotting methods allowed them to be easily reused across many validations, and the definition of measured variables by their relationships to lower-level variables rather than the procedures measuring them allows their reuse across many models and validations. 

% TODO: some redundant sentences here
While our decomposition of the model-side code into various filters is not, strictly speaking, a part of the pyRMV framework (the model code can be organized any way the modeler prefers), it integrates more easily with our data-based model interfaces than it would with a procedural interface. The fact that the model provides predicted properties instead of implementing experiemental operations gives much more flexibility to its implementation and thereby enables this decomposition. Implementing such a decomposition would be much more challenging if not impossible in an operationalist framework. The way in which the model class so effectively encourages code reuse therefore indirectly demonstrates the benefits of the pyRMV framework.

A major strength of this framework lies in the simplicity of defining validations: due to the reusability of statistical and plotting methods, a validation can usually be defined simply by standardizing its data. This will allow the library of available validations to grow quickly and increase the benefit to modelers. The simplicity of defining validations is an especially important strength when we consider the position of an experimentalist wishing to assess the degree to which their novel finding contradicts established models, as the experimental scientist is likely to be less well versed in programming than a computational scientist is. 

In terms of both model and validation code, the usability and simplicity of pyRMV is made possible by its relatively small number of abstractions. Those abstractions that are present, such as plotting, statistical tests, models, validations, and data tables will already be familiar to contributors. This also means that model code deals either with matters of the model's implementation or with concepts in the scientific domain under study, both of which should be familiar to the modeler. This minimal abstraction is made possible by eschewing inheritance hierarchies for composition \parencite{gamma_design_1994} and complex data structures for tables. 

Due to the flexibility provided by its interfaces we expect pyRMV to generalize well both to very different models and to very different experiments than we have used it with here. However, we cannot be certain of this until it has been applied in practice to a wide variety of models and experiments.

In addition to these strengths, the examples illustrated and clarified several challenges faced in systematic model validation.
Future versions of pyRMV can address these challenges better.

\subsection{Limitations}
\label{sec:limitations}
Many of the challenges and limitations we faced were related to the problem of holism in scientific testing; any interpretation of an experiment and its results is conditioned on a framework of background knowledge about the system under study. Challenges \ref{ch:describe-experiment}, \ref{ch:theory-ladeness} and \ref{ch:poorly-defined} are all reflections of this problem.  Challenges \ref{ch:model-constraint} and \ref{ch:scope-issues} are instead implementational in nature.

\subsubsection{Challenge \ref{ch:describe-experiment}, experimental biases}
In the firing rate example, we saw how it is not always possible to remove the experimental operations from consideration in the validation process. 

With an operationalist validation approach it is made explicit what is required for a model to accurately reproduce the measured quantity. With our realist approach, in contrast, the experimental operations are abstracted to additional variables for the model to optionally take into account. Which experimental variables are of interest will depend on the model's assumptions.
Communicating the details of the experiment in a clear and unambiguous manner is more difficult with our method, as the variables are divorced from the code that gives them meaning; we must rely instead on the description on the Term object.

However, even with an operationalist approach it would be impractical to describe the experiment exhaustively. As we have argued, the set of instructions for modeling the experiment will include simplifications and abstractions which embed assumptions about the target system in the validation.

The advantage of our approach is that it leaves any such simplifications and assumptions up to the model in question. It does not exclude models incapable of representing the experimental procedure (design goal 4), and does not \emph{require} the implementation of that capacity in models where it is possible but not expedient (supporting design goal 1).

% TODO: here relate to how we address multiplicity of orientation selectivity
Due to this flexibility, we could create one or more representations of the experimental procedure and provide these as reusable modules for models to use, thereby adding meaning to the passed experimental variables and combining the merits of operationalist and realist validation. If designed to operate on predicted properties, such a model of the experiment could be fairly broadly applicable. What is then effectively validated is the union of a given model of the target system and a model of the experiment.

\subsubsection{Challenge \ref{ch:theory-ladeness}, theory-laden variable interperetations}

In the examples of relative innervation strength and fraction excitation per connection we showed how the choice of variable to validate with a given experiment is informed by beliefs about the target system. We also showed how the beliefs embedded in the model, experiment, and validation may differ. In the case of fraction excitation per connection, this could not be resolved by passing the method of measuring convergence to the model, as this would have prevented the statistical test.

For relative innervation strength the issue could be reconciled by choosing a specific form of synaptic strength, unitary PSC amplitude. However, had we aimed to validate a model without short-term plasticity we would likely have created a more ambiguous validation. The same is true of the fraction excitation per connection validation - it would have been easy to unknowingly embed additional assumptions in the validation. This raises the question of what assumptions of our model we have unthinkingly embedded in other validation decisions. 

% note: the reason we couldn't do convergence=80 and pass the method of computation is that the statistical test would no longer make sense: sampling inferred convergence from target neurons and looking for an average wouldn't be the same as sampling fraction excitation.

In reality, we likely will not realize what possibly mistaken assumptions are embedded in such decisions until someone attempts to validate a model which disputes those assumptions. At this point, the validation could be changed, but unlike adding additional experimental variables, changing the measured variable of a validation breaks its interface with models. 
The modified validation could be a separate validation, but this leads to the fragmentation of validation for different modeling approaches, while one of the advantages of systematic validation is that it permits the comparison of different approaches by the same metrics. In time, it would be best if the disagreement regarding the interperetation of experiments is resolved and consensus established, at which point we will need a means of adding a deprecation notice to the 'losing' form of the validation, directing users of it to the 'winning' form.

%\ref{ch:theory-ladeness} choosing fraction excitation over lower property, choosing relative innervation strength rather than pathway PSC amp.

\subsubsection{Challenge \ref{ch:model-constraint}, generalization of model constraints}

In both the firing rate and orientation selectivity validations we encountered ways in which the computational limitations of the current model informed the construction of a validation intended for any model. If the raw data were not available, we would have been forced to use the experimental data as-is and have the model attempt to predict them with a reduced number and duration of simulations. We could have chosen to do this in any case, but if the model fails to predict the experiment it would be unclear whether this is because of a flaw in the model itself, or in the imputation of its reduced simulations onto the full experimental set. 

Ultimately, validations should run as quickly as possible. Therefore it is preferable to format the experimental data so that it can be replicated with less simulation time, so long as the experiment remains sufficient to assess the validated property, even for more computationally efficient models. 

This neatly addresses the challenge when applied to the duration of simulations, but it does not apply neatly to the case of limiting the stimulus set in the orientation selectivity validation. Limiting the set of stimuli omits details which are likely to be relevant to the phenomenon being studied. 

This could instead be framed as a matter of scope. If simulating responses to all stimuli is impractical for the model in question, then reproducing those details is out of scope for it. In this case, we would have multiple different validations: some using a subset of the stimulus set, and some using the full stimulus set to capture more detail.

%\ref{ch:model-constraint} firing rates/osi simulation duration, osi set of stimuli
\subsubsection{Challenge \ref{ch:scope-issues}, incomplete scope overlaps}

The validations for firing rate and orientation selectivity included data for some brain regions not included in our model. As a result, the first versions of our validation code ran successfully when the other brain regions were omitted from the validation but crashed when the full validation was used.
By modifying the statistical analysis, plotting, and model code we were able to make the validation run for the full data as well. 

For systematic validation to work it needs to be possible for the validator to create a new validation and then run it on models which have never run that validation before, provided they predict the required variables, without needing to modify their code. If the scope of a validation has overlap with the scope of a model, then it is useful for that model, and should run successfully on it. However, at present, the model may crash when presented with a variable combination outside of its scope.

The burden is presently on the modelers to design their model interface classes carefully. There are a few ways this could be addressed, such as allowing modelers to specify their model's scope with decorators or requesting measurements one-by-one and interpereting any error message as NaN (with a suitable warning). The simplest solution is for models to trim the experimental variables to exclude values which are outside of their scopes, and return predictions only for the relevant values (a tool to facilitate this could be included in future versions of the framework). The Analysis class could then restrict analysis only to those values which at least one model predicted.

%\ref{ch:scope-issues} predict firing rate for only subset of regions

\subsubsection{Challenge \ref{ch:poorly-defined}, vaguely defined properties}

OSI can be calculated in a variety of subtly different ways, as well as being applied to different signals.

There are two ways for our conceptual approach to handle this: by treating each variation of OSI as its own property, or by treating them as the same property and creating additional variables to describe different variations of it. Each of these has its issues in terms of clarity and terminology bloat.

Another possiblity would be to measure firing rates rather than orientation selectivity, and calculate OSI in the specific way of the experiment within the statistical methods. This would preclude the application of analyses to models which represent orientation selectivity directly, without modeling the underlying activity. It is hard to imagine why any model would choose to do so, but if we can avoid it, it is better not to let our current imagination constrain future models. 

In this case we consider the advantages working on firing rates directly to outweigh this nebulous concern. Besides allowing specific orientation selectivity indices to be defined for different validations without additional terminology or model-side code, it would also simplify statistical analysis.  The validation could shuffle firing rates internally and compare the shuffled model to the original, without a separate control model needing to be passed to the validation.

% TODO: here, we need to generalize our solution so that it automates the application of the appropriate OSI definition.
% define functions to calculate different OSIs from lower-level properties
% avoid terminology bloat by referencing these functions in the validation instead of Term objects for the different OSIS
% effectively implements limited operationalism 
This decision would echo that made in the fraction excitation per connection validation, where the relevant statistical analysis could only be made my measuring an underlying property. These two examples make it clear that there are circumstances where it is better to compromise the principle of using the highest-level property. This shows a practical limitation of a purely realist approach, and suggests it may be best to implement a compromise between operationalist and realist methods. In particular, it suggests that realism should be selective, and the selection of properties to reify should depend on the ambiguity or concreteness of the property in question.

%\ref{ch:poorly-defined} OSI can be defined many ways and applied to many things, operationalist method would not require definition of new methods per se, while ours requires either new methods, or additional parameter handling

\subsubsection{Challenge \ref{ch:model-inefficient}, slowness of generalized implementations}

In the connection probability and orientation selectivity examples it was shown that the initial implementation of measurement code can often be dramatically less efficient than of a standalone script. Caching mostly mitigated this, but the effort required to get a working first version of the model-side validation code was nonetheless noticably greater than for a standalone validation, and the result was still noticably slower. However, much of this effort would not need to be repeated for subsequent related validatins, which were relatively straightforward to implement (design goals 1, 2).

We are considering requesting measurements on a row-by-row basis rather than all at once (see below). In such a case, optimizing performance through grouping and vectorization would be precluded entirely. This can be offset by parallelization.

\subsubsection{Other limitations}

%relationship of describe-experiment and theory-ladenness
In experiments, the many stimuli are shown to the subject in a single session, and it is entirely conceivable that preceding stimuli influence responses to succeeding stimuli. The current version of our framework does not support modeling such effects, as each stimulus is considered in isolation and it is not relayed to the model which other stimuli were shown alongside it in the experiment. We currently have no plan or suggestion to correct this, but are aware that it may be a consideration in the future. An operationalist approach to the experiment, in which the full experiment is simulated and a variety of different validations performed on the different aspects of it solves this problem most naturally, although it heavily constrains which models it can be applied to in the process. It remains an open question how best to account for this with a realist approach.

The visual stimuli in the validations we used so far are described parametrically. As pointed out by \cite{schrimpf_integrative_2020}, the ability to broadly validate modern visual neuroscience models is contingent on their ability to respond to arbitrary videos. With parametric stimuli, each model must implement some code for every type of stimulus, while if a video is provided the model can use this directly. It is important that, alongside metadata like stimulus type and orientation, videos are provided to the model as a part of the visual stimuli.
This same reasoning can be applied to other stimuli, such as current pulses in electrophysiology experiments: providing a time-series of the stimulus allows a model to simulate responses to any stimulus without needing separate code for each protocol.

At present, we rely only on the informal specifications for data types written in the terminology module. This makes testing for the correctness of validation data more challenging. As the size of the library and the user base grows, we have to adopt formal table schema validation approaches that verify the values in a table are appropriate. It may be necessary to adjust some aspects of our approach to facilitate this. 

In creating a validation, a user of the framework may create simple, specialized statistical analysis and plotting methods. In merging their validation into the library, during the process of code review and refactoring, they can be incentivized to rewrite appropriate parts in a reusable manner. However, the same cannot be said for code implemented on the model-side which may be useful to other modelers. If we had an operationalist approach that kept more of the experiment representation on the validation side, it would be easier to encourage generalizability.

Currently, a fairly large proportion of the code in the statistical and plotting methods involves the grouping of data by dependent and compared variables. Creating higher-level statistical and plotting abstractions could greatly simplify the process of creating reusable statistical and plotting methods.

While the framework supports validations with multiple measured variables and a validation doing so has been implemented, this currently relies on the model providing an appropriate index for both measurements, which is not otherwise enforced. Ensuring that models use consistent indexing for different measurement types, for instance throygh schema validation, would help ensure validations with multiple measured variables work consistently for different models.

At present, the domain-independent parts of this framework are in the same repository as the neuroscience parts. For the framework to be applied to other domains these would first need to be separated. As the framework grows, it may be necessary to divide the neuroscience repository further. However, implementing disciplinary boundaries into the code in this way may have detrimental effects that undermine the purpose of a systematic validation library such as this by applying different standards and bodies of evidence to models with overlapping but non-equal scopes. Fragmentation is best avoided, so that different modeling approaches can still be meaningfully compared.

%As the size of the library of validations grows, it may become necessary to add searching and filtering methods need methods for filtering validations as size of library grows

%also discuss the need for lazy data loading

%discuss the inconsistency between sample-based methods for some measurements, and not for connection probability, fraction innervated.
%discuss merits and failings of probability-distribution based evaluation approach

%discuss incentivizing generalized methods through contributor guidelines and review processes, while permitting quick-and-dirty development of an initial version of the validation.

%\todo{consider saying something about how LLM-assisted coding can benefit this approach?}

%(the disadvantage of viewing, e.g. cossell and lee as measuring the same things is that it embeds assumptions about the accuracy of their methods into the validation, which a model may dispute. However, the flexibility of our approach allows space for such dispute, s.t. a model may include reproduction of the errors and biases of the experimental approach).

Software development constantly presents new challenges. The challenges we discussed here are only those which we could clearly link to methodological issues and articulate as such. It should not be taken as an exhaustive list of the difficulties in creating such a framework.

\subsection{Future directions}
\subsubsection{Batch-processing v.s. row-by-row measurements}
\label{sec:batch-process}
% add this later
In the current version of the framework, we request all measurements for a validation from the model at once. This batch-processing provides additional flexibility to the model, allowing it to use information about the experimental variables to optimize the computation and determine how to parallelize any measurement procedures. However, several downsides of this decision were made clear.

Firstly, in the creation of reusable methods  for different connectivity validations  we found that in requesting the underlying \verb+pair_weight+ variable in one batch, a large amount of memory ($>$100GB) was required to contain the resulting dataframe. This means that while a model may benefit in terms of speed by batch-processing, they are forced to use a lot of memory if they make use of the reusable methods.

Secondly, we found that the code required for the batch-processing interface was more complex and less expressive. This is because each method required additional lines for iterating, appending or grouping by experimental variables. If the measurements were requested from the model one-by-one the model code would have been simpler. For most measurements, we chose to iterate through the measurements anyway for this reason.

It may therefore be preferable to request measurements one at a time, doing so asynchronously to allow for parallelization.

%\ref{ch:model-inefficient} connectivity validations batch processed, similar for OSI

\subsubsection{Handling subject-to-subject variation}

In every validation we performed we validated the distribution of the measured variable in the model against that in the experimental data. However, our model represents the distribution of the variable as observed from a single, arbitrary mouse. Some experiments reported data from a single mouse, and some from multiple mice. A mismatch between the distribution of the model, and the distribution from a single mouse does not neccessarily indicate that the model's representation of that property is implausible. Similarly, if a model does not reproduce the degree of variation in a variable observed across multiple mice, it does not mean that it does not represent a plausible mouse.

We can update the reusable statistical methods to distinguish subject-to-subject variation from intrinsic variation when a SUBJECT variable is present, then all validations making use of these methods benefit automatically.

To compare two datasets where at least one involves multiple subjects we could follow the approach used by \cite{billeh_systematic_2020}: The distributions of each subject are compared to those of each other subject, resulting in two distributions of test statistics: experiment-to-experiment and experiment-to-model, after which we can test that these two distributions are the same.

\subsubsection{Relaxing the highest-level feature constraint}

The examples of orientation selectivity, relative innervation strength, and fraction excitation per connection show additional downsides of only directly requiring the validated feature in a validation. Several limitations may be resolved by softening this requirement. 
In this case, we can expect from the model only well-defined properties (such as firing rates), while more complex features (such as orientation selectivity) can be defined programatically.
These can therefore be defined in different ways for different validations.

The main criterion in evaluating whether to validate a variable is how concrete it is.
For example, synaptic strength is a concept that can mean many things, and it is better to refer to a specific form, rather than the general concept.
The downside of doing so is that a model must implement a different method for each form, but  as the translation between them is often trivial this may not be a major issue (for example, multiplying synaptic weight by release probability to go from post-synaptic density area to post-synaptic current).

\subsubsection{Expanding scope}

%While the software package is aimed specifically at network models in neuroscience, the principles used could be applied to any domain of computational science.

The current implementation of this framework focuses on validation of models with respect to a class of target systems that it represents by means of comparison to experimental observations. Importantly, this includes validating a model's parameters and internal states in addition to its outputs, thereby checking not only accuracy but verisimilitude. Certain forms of validation are excluded by this definition, but we emphasize that our approach can be expanded to accommodate them.

Firstly, some models are intended to accurately model specific instances of a system rather than a class  of systems or an arbitrary instance thereof. Assessing the quality of the fit of these models to their target instances is out of scope for this article, though we explain below how pyRMV might be extended for this case. Models which aim to reproduce specific instances can still benefit from checking the correspondence with the class of systems the target belongs to, for instance, through prior predictive checks \parencite{gelman_bayesian_2020,winter_illustrating_2023}, and therefore can still benefit from the current version of this framework. This way it can be more accurately evaluated how well the model reproduces the processes of the target system, and a larger pool of data can be used to constrain the model and prevent over-fitting
.
To extend this framework to assess the fit to a specific instance of a system, one can simply include information about the specific instance in the experimental variables. In many cases, models are calibrated to fit the instance they represent. In this case some output variables are measured from the target system, the model fit to a subset of them, and validated on a different subset. The validation procedure can then provide the training data subset in the experimental variables as a part of the data on the basis of which the model's predictions are made.

Model verification, that is, checking whether the model is correctly implemented \parencite{oberkampf_verification_2010,graebner_how_2018}, is also excluded from our scope. Due to focusing on comparisons with experimental data, our scope also excludes forms of validation in which it is checked whether well-known numerical laws emerge from the model. In principle this can be created in the present framework by initializing an Analysis without the observed variable in the observations dataframe, we have not yet done so and therefore cannot highlight any particular challenges involved. It should be noted that since model verification is specific to any one conceptual model, the benefits of universalizing it are therefore more limited than with validation.

The current implementation of this framework was designed for models which provide instances of the predicted values for a property, such as \cite{billeh_systematic_2020} or \cite{dictus_we_2026}. For example, when the model predicts the firing rates of neurons, it provides one firing rate for each of the neurons it models. Not all models would represent a predicted property with a series of samples. Some models provide a single expectation or most likely value. Other models may provide the parameters of a distribution, or a lower and upper bound on a predicted feature. In order to keep statistical tests assessing model quality consistent, models of these kinds are not explicitly supported by the current version, although some can be rendered compatible with it through sampling. Some of such models may require specific statistical methods for comparison with the experiments, and supporting them may therefore be challenging future work. 

While all validations in the framework are designed to seamlessly enable model-to-model comparisons, we have not implemented any approach to pooling results between validations or ranking models. In order to rank models with respect to a given validation, the p-value of the statistical tests used to assess the pass/fail state can be used as a metric. Alternatively, different statistical analyses based on other metrics can be created for the purpose of ranking models. 
Ranking models by pooling the results of multiple validations, as some frameworks attempt to do \parencite{schrimpf_integrative_2020,gerkin_towards_2018}, is more complicated, and we do not currently have a means to meaningfully do so for models with overlapping but non-equal scopes. However, by considering only those validations that both models support, their suitability for particular purposes could be evaluated, which future versions of pyRMV may facilitate.

It should be noted that we do not distinguish within this framework whether we are validating the quality of fit (descriptive validation) or predictive capacity of the model (predictive validation). A validation is blind to whether its data were used in the construction, calibration, or selection of a model. These distinctions are important when assessing whether a model can be relied upon. Developing methods to efficiently document this information and incorporate it in the validation and comparison of models is important future work. Two possible ways of addressing this would be by either relying on modelers to specify which data informed their model, or to use the timings of models and experiments to determine which models could have used a given dataset.

\subsection{Future prospects and impact}

%TODO REVISIT AND PUT SOME POINTS FROM THE PROPOSAL IN HERE

Addressing the challenges involved in systematic validation can accelerate computational science, and through it, all fields of science in which it is applied. Our contribution can begin to realize these benefits in visual and simulation neuroscience.

The first way in which this can be done is by accelerating the refinement of existing models. Through implementation-independent interfaces, existing models can be validated anew with little effort when they are improved, while permitting a great deal of implementational flexibility. Expanding the range of data with which a model is validated is also made substantially easier by the code reuse strategies developed here.

The second way is by making it easier to validate completely new models. With a large library of validations, the data curation steps could largely be skipped. With only a small codebase a new model could be valiidated against a wide variety of data and immediately compete with more established models. To realize this goal, maximum implementational flexibility is neccessary, which our framework aims for.

The third way is by allowing modelers and experimentalists to compare many models without needing to dive into their internals. After importing the relevant model classes they can be queried using a common interface. This also allows experimentalists to quickly verify which well-established models are consistent or inconsistent with their new observations, provided the terminology to describe those observations has already been defined. As the size of the library of validations grows it will become increasingly likely that the needed terminology is defined. This has potentially far-reaching implications.

\cite{churchland_conceptual_2016} argue that if a global understanding of the brain is to emerge, it will be as a ``patchwork quilt'' of different models and approaches. This echoes pluralistic descriptions of scientific enterprise from philosophers such as Lakatos and Feyerabend \parencite{feyerabend_against_2010,lakatos_criticism_1969}, in which different paradigms or research programmes exist in parallel. Modern multi-scale simulations show how different paradigms can be used in concert and complement one another \parencite[c. 1,2]{winsberg_science_2010}, and so need to be validated together as well as individually. The specialization central to Churchland and Abbott's description is a means of handling the immense complexity of the field: breaking it down into more manageable sub-fields and subsequently combining insights from them.

This is analogous to a key software engineering strategy for handling complexity: modularization. The complex problem is broken down into components and each of those components isolated from the others. Identifying the boundaries of different modules and defining the interfaces between them is critical for this. In this way it is not neccessary to know the internals of one module when working on another, only its interface: what it does, not how it does it.
Equally critical is automated and systematic testing: it is essential that when a module is modified it continues to do everything it previously could, or else a change in one module may break the functionality of any other module using it. Other modules need to be able to rely on a given module without fully understanding its internals, and testing helps to guarantee that they can.

This analogy illustrates why validation is an essential part of the stitching of the aforementioned ``quilt''. By defining the scopes of validity of different models and testing them thoroughly, we empower the use and re-use of models in the neuroscientific community.

If we are thoughtful in our design, these same benefits can be applied to any other field.
Systematic validation will be indispensible for modeling deeply complex systems such as organs, organisms, and economies.
It is necessary in order to progressively expand the scope and accuracy of such models, as well as to smoothly switch tracks to new approaches.
Likewise the standardization of model interfaces will allow multiple domains of knowledge to be combined, and the complexity of scientific domains more effectively managed.
For this reason, the construction of systematic, standardized validation may be the most important challenge in computational science.
PyRMV and the operationalist/realist distinction represent crucial steps towards addressing this.
Many of the relevant issues have yet to be explored, and solving them will require participants with expertise in the scientific domains of interest, software engineering, design, and philosophy of science.

\section{Author contributions}

Hugo designed and implemented the framework, and wrote the article. Eduard constructed a validation and provided feedback on the framework and code. Armando and Henry provided guidance, discussion, and feedback on the article.

% Standardized, systematic validation addresses several related problems:
% Fistly the problem of comparison: comparing conceptually distinct models in their realism, strengths and limitations
% Secondly, the problem of completeness: comprehensively validating any single model- infeasible if validation has to be done from scratch all the time
% Thirdly, the problem of interoperatbility - it is not a complete solution for this, but the model abstraction created is equally essential for this

\newpage

\printbibliography

\end{document}